\documentclass[pra,pra,twocolumn]{revtex4}
\usepackage{graphicx}

\newcommand{\doubleslash}{\rm{/\!\!/}}

\begin{document}
\title{Thermo-modulational interband susceptibility and ultrafast temporal dynamics in nonlinear gold-based plasmonic devices}
\author{Andrea Marini$^1$, Matteo Conforti$^2$, Giuseppe Della Valle$^3$, Ho Wai Lee$^1$, Truong X. Tran$^1$, Wonkeun Chang$^1$, Markus A. Schmidt$^1$, Stefano Longhi$^3$, Philip St.J. Russell$^1$ and Fabio Biancalana$^1$}
\affiliation{$^1$Max Planck Institute for the Science of Light, Guenther-Scharowsky-Stra\ss e 1, 91058 Erlangen, Germany \\$^2$CNISM and Dip. di Ing. dell'Informazione, Universit\'a di Brescia, Via Branze 38, 25123 Brescia, Italy \\
$^3$Dip. di Fisica and IFN-CNR, Politecnico di Milano, Piazza L. da Vinci 32, I-20133 Milano, Italy}
\date{\today}
\begin{abstract}
Starting from first principles, we theoretically model the nonlinear temporal dynamics of gold-based plasmonic devices resulting from the heating of their metallic components. At optical frequencies, the gold susceptibility is determined by the interband transitions around the $X,L$ points in the first Brillouin zone and thermo-modulational effects ensue from Fermi smearing of the electronic energy distribution in the conduction band. As a consequence of light-induced heating of the conduction electrons, the optical susceptibility becomes nonlinear. In this paper we describe, for the first time to our knowledge, the effects of the thermo-modulational nonlinearity of gold on the propagation of surface plasmon polaritons guided on gold nanowires. We introduce a novel nonlinear Schr\"odinger-like equation to describe pulse propagation in such nanowires, and we predict the appearance an intense spectral red-shift caused by the delayed thermal response.
\end{abstract}

\maketitle

\section{Introduction and motivations}

The design and development of subwavelength photonic devices with metallic components has become a subject of intense research in the last decade. This trend is justified by the need for compact high-performance optical devices and is mainly driven by the enormous technological improvement in nano-fabrication techniques.

These state-of-the-art manufacturing tools for metallic nano-circuits have made it possible to design and engineer the effective optical properties of artificial materials, commonly known as {\it metamaterials} \cite{Engheta_Book,Ramakrishna_Book}. In these synthetic materials, the propagation of light is strongly influenced by the geometric properties of the embedded metallic nano-circuits. In particular, the metallic nano-structures can be tailored in such a way that the effective refractive index becomes negative \cite{Veselago_1968,Shelby_Science_2001,Smith_Science_2004,Zhang_PRL_2005,Shalaev_OL_2005,Dolling_Science_2006,Dolling_OL_2006,Dolling_OL_2007,Yang_PRL_2011}. Negative index materials (NIMs) are potentially important in superlensing \cite{Pendry2000,Shalaev2007,Kawata_NatPhot_2009} and cloaking applications \cite{Schurig_Science,Cai_NatPhot_2007,Valentine_NatMat_2009}. Novel physical mechanisms occur in anisotropic metamaterials, which under some circumstances can exhibit hyperbolic dispersion \cite{Jacob_APB_2010,Noginov_OL_2010,Smolyaninov_PRL_2011,Kidwai_OL_2011,Soukoulis_NatPhot_2011}. In the nonlinear regime, metamaterials have been studied for harmonic generation \cite{Shadrivov_JOSAB_2006,Klein_Science_2006,Shadrivov_APL_2008,Vincenti_PRA_2011,Ciattoni_PRA_2012}, soliton propagation \cite{Shadrivov_OE_2005,Marklunda_PLA_2005,Shadrivov_JOA_2005,Liu2007,Rizza_PRA_2011} and optical modulation and switching \cite{Husakou_PRL_2007,Min_OL_2008,Ciattoni_OL_2010,Ciattoni_PRA_2011,Gong_OE_2011}.

The unusual properties of metamaterials arise from the fact that the metallic nano-circuits are much smaller than the wavelength of light, resulting in a space-averaged macroscopic dielectric response. Conversely, in the case where the optical wavelength is comparable with the dimensions of the metallic sub-structures, the light feels the geometric details and plasmon polariton modes are excited. Surface plasmon polaritons (SPPs) are electromagnetic waves propagating on metallic surfaces \cite{Maier_Book}. They constitute the best candidates for manipulating light on the nanoscale and for the development of subwavelength all-optical devices \cite{Brongersma_Book,Bozhevolnyi_Book,Ozbay2006,Schmidt_PRB_2008}. In particular, plasmonic waveguides have important applications as optical interconnects in highly-integrated optoelectronic devices \cite{Bozhevolnyi2006}. Other relevant applications of SPPs are found in medicine \cite{Gobin_NanoLett_2007}, sensing \cite{Homola_Book,Vedantam_NanoLett_2009,LeRu2009} and nano-lasers \cite{Bergman2003,Protsenko_PRA_2005,Stockman_NatPhot_2008,Zheludev2008,Oulton_NatLett_2009,Marini2009,Gather_Nature_2010}. The nonlinear properties of SPPs can be used for second harmonic generation (SHG) \cite{Zayats_Book}, active control \cite{Wurtz2006,MacDonald2009} and nanofocusing \cite{Durach_NanoLett_2007,Davoyan_PRL_2010}. Nonlinear self-action can be exploited for manipulating transverse spatial diffraction by self-focusing \cite{Davoyan2009} and for the formation of plasmon-solitons \cite{Feigenbaum2007,Marini_PRA,Marini_OE_2011}. Fundamental studies of metamaterials and SPPs are closely related. Indeed, relevant phenomena occurring in metamaterials are observed also in plasmonics, e.g. negative refraction \cite{Dionne_OE_2008,Liu_OE_2008}, anomalous diffraction \cite{ConfortiGuasoni2008,DellaValle_OL_2010} and electromagnetic cloaking \cite{Alu_JOA_2008,Edwards_PRL_2009,Alu_PRL_2009}. In both fields, the innovative step is the use of nanostructured metals for manipulating light.

In most of the nonlinear studies reported above the optical response of metals is assumed to be  {\it linear}, while the nonlinearity originates from the dielectric medium; however, experimentalists know well that the Kerr nonlinearity of metals can be enormous. Experimental results indicate strong third-order nonlinear susceptibilities that vary by several orders of magnitude, with values of $\chi_3^m$ that vary between $10^{-14}$ and $10^{-18} m^2/V^2$ \cite{Ricard_OL_1985,Yang_OptMat_2004,Uchida_JOSAB_1994,Smith_JAP_1999,Lee_JOSAB_2006,Rotenberg_PRB_2007} and that are much bigger than the third order susceptibility of bulk silica ($\chi_3^{Si} \approx 10^{-22} m^2/V^2$). Recently, the nonlocal ponderomotive nonlinearity for a plasma of free electrons has been proposed as a possible model for the interpretation of experimental results \cite{Ginzburg_OL_2010,Davoyan_PLA_2011}. The predicted value for the ponderomotive third-order susceptibility at optical frequencies ($\chi_3 \approx 10^{-20} m^2/V^2$) is however insufficient to explain the experimental findings. In addition, the spectral dependence of the ponderomotive nonlinearity ($\chi_3 \propto 1/\omega^4$) does not fit with the enormous spectral variation (by several orders of magnitude) observed in the measurements, suggesting that the basic nonlinear mechanism for metals is {\it resonant}. Theoretical and experimental confirmation of this hypothesis is to be found in the results of Rosei, Guerrisi {\em et al.} on the thermo-modulational reflection spectra of thin films of noble metals \cite{Rosei_PRB_1972,Rosei_SurfSci_1973,Rosei_PRB_1974,Rosei_PRB_1974_bis,Guerrisi_PRB_1975}. In their work, the authors theoretically predict and experimentally observe a strong modulation in the reflection spectrum due to light-induced heating. They demonstrate that the temperature change smears out the energy distribution of the conduction electrons, affecting the resonant interband absorption and hence the dielectric susceptibility. This process is intrinsically nonlinear, since the temperature change modulating the dielectric response depends on the optical power. Subsequent pump-probe experiments in thin films \cite{Sun_PRB_1994,Groeneveld_PRB_1995,Hohlfeld_ChemPhys_2000} and nanoparticles \cite{DelFatti_PRB_2000,Baida_PRL_2011} have confirmed the initial results of Rosei and Guerrisi. Theoretical and experimental investigations on the temporal dynamics of the system clearly indicate that the nonlinear response of metals is characterized by a delayed mechanism \cite{Sun_PRB_1994,Carpene_PRB_2006}, as is typical for thermal nonlinearities \cite{Boyd2003}.

Very recently, a complete analysis of the nonlinear optical response of noble metals, leading to the first theoretical derivation of a consistent model for the third-order nonlinear susceptibility of gold, was reported \cite{Conforti_PRB_2012}. Although experiments in thin films have been satisfactorily explained \cite{Conforti_PRB_2012}, a theoretical description of the thermo-modulational interband nonlinearity for ultrashort optical pulses propagating in plasmonic waveguides is still missing.

In this manuscript we derive the thermo-modulational nonlinear susceptibility reported in \cite{Conforti_PRB_2012}, starting from the band structure of gold, and describe its effect on SPPs propagating in a gold nanowire surrounded by silica glass. The paper is organized as follows. In section I we describe the optical properties of gold, the interband transitions, their effect on the dielectric susceptibility and its temperature dependence. In section II we model the temporal dynamics of the electrons through the two-temperature model (TTM), deriving the characteristic temporal response function. Finally, in section III we model the propagation of SPPs along a gold nanowire by introducing a novel nonlinear Schr\"odinger-like equation and predicting for the first time to our knowledge a strong red-shift caused by the thermo-modulational nonlinearity of gold.

\section{Optical properties of gold}

The nonresonant optical properties of metals can be described through the free-electron model, where electrons are considered as free charges moving in response to an optical field ${\mathrm Re}[\vec{E}_0e^{-i\omega t}]$ oscillating at angular frequency $\omega$. In this model, the dielectric response of the plasma can be derived directly from the non-relativistic single-particle equation of motion \cite{Ashcroft_Book}:
\begin{equation}
\epsilon_{intra} (\omega) = 1 - \frac{\omega_p^2}{\omega^2 + i \gamma \omega}, \label{Free_Electron_Model_Susceptibility}
\end{equation}
where $\omega_p=\sqrt{n e^2/\epsilon_0 m_e}$ is the plasma frequency, $\epsilon_0$ is the vacuum permittivity, $n$ is the electron number density, $e,m_e$ are the electron charge and mass and $\gamma$ is a characteristic frequency accounting for electron-electron collisions. This model is justified by the fact that for metals the Fermi energy lies within the conduction band and many accessible states exist for the electrons. From a quantum perspective, free-electron motion only accounts for {\it intraband} transitions.

\begin{figure}
\centering
\begin{center}
\includegraphics[width=0.45\textwidth]{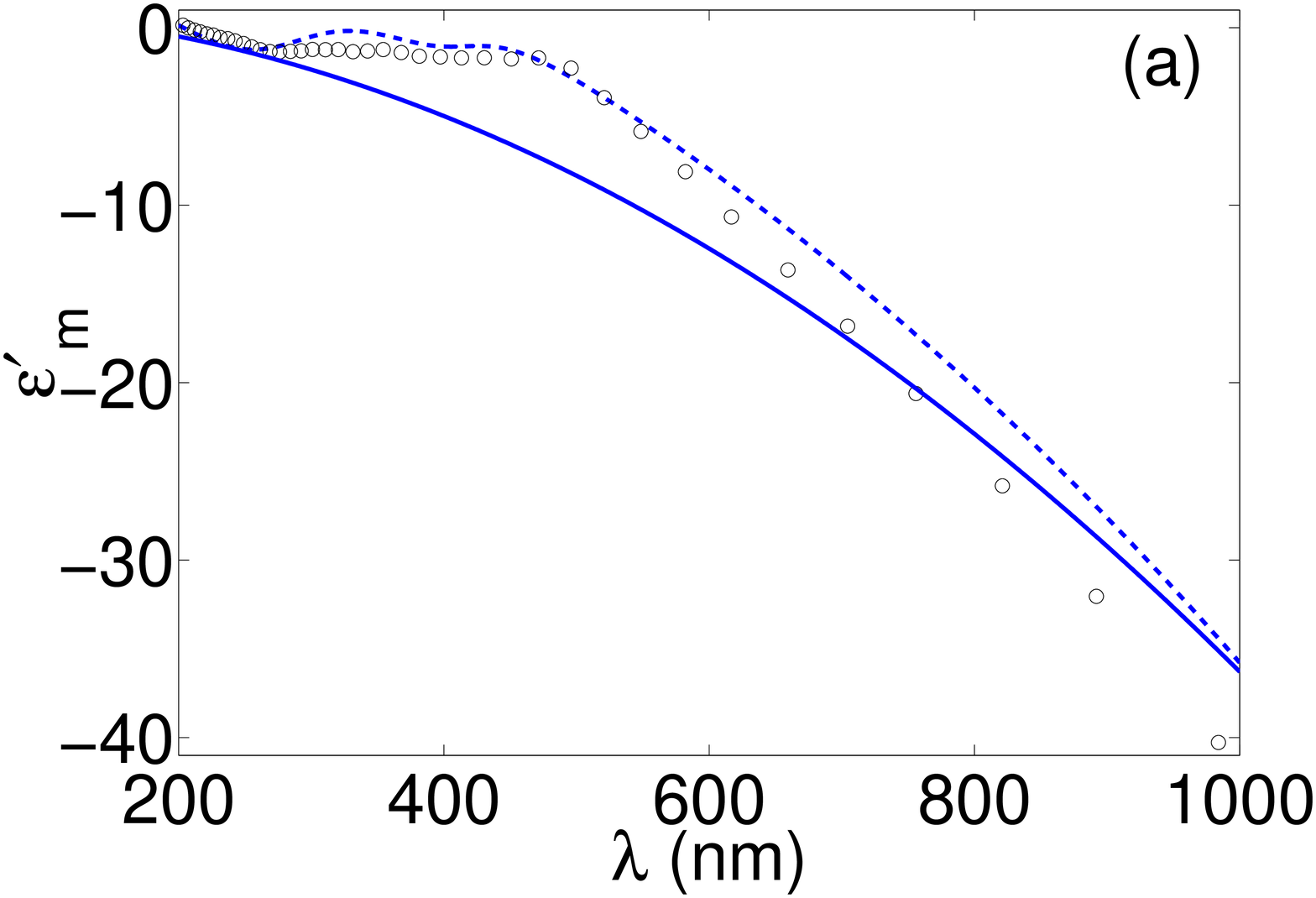}
\includegraphics[width=0.45\textwidth]{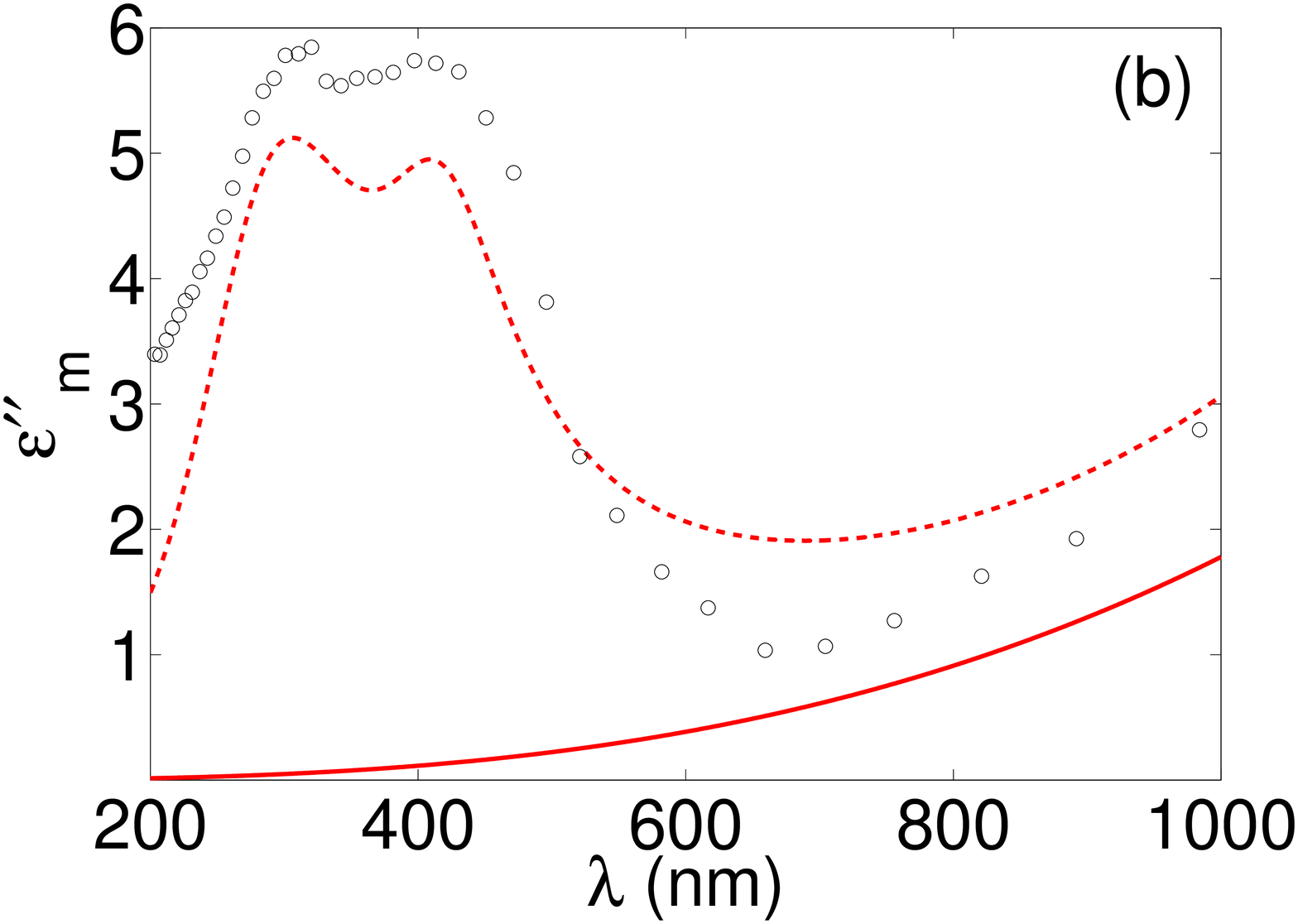}
\caption{(a) Real and (b) imaginary parts of the dielectric constant of gold. The full lines represent the free-electron prediction $\epsilon_{intra}$, while the dashed lines correspond to a fit to the experimental data in Ref. \cite{Ung2007}. The open circles represent the experimental data points of Johnson and Christy \cite{Johnson1972}.}
\label{Gold_Diel_Const}
\end{center}
\end{figure}

For wavelengths in the far-infrared, the free-electron model provides very good quantitative agreement with experimental data for all noble metals \cite{Ung2007}. For the special case of silver, the nonresonant model also works well at optical frequencies. In contrast, gold and copper susceptibilities have properties that are more involved and the nonresonant model is not appropriate at optical frequencies or in the near infrared. Indeed, {\it interband} transitions between the $d$-band and the conduction band become more important and cannot be neglected if one wants to model accurately the optical response of such metals \cite{Ashcroft_Book}. The presence of interband transitions enriches the variety of physical processes occurring in metals and lies behind the strong temperature dependence of the dielectric susceptibility. In what follows, we focus on the particular case of gold and its dielectric properties. In Figs. \ref{Gold_Diel_Const}(a,b) the real and imaginary parts of the dielectric constant of gold are plotted. Full lines represent the free-electron prediction $\epsilon_{intra}$, while dashed lines correspond to a fit to the experimental data in Ref. \cite{Ung2007}. The open circles represent the experimental data points of Johnson and Christy \cite{Johnson1972}. For the free-electron calculations we used the parameters $\omega_p = 1.1515 \times 10^{16} rad/sec, \gamma =  8.9890 \times 10^{13} sec^{-1}$, obtained by fitting $\epsilon''_{intra}(\omega)$ to the experimental data for long wavelengths in the far-infrared. Note that at optical frequencies the measured dielectric susceptibility deviates significantly from the predictions of the free-electron model as a consequence of two intense absorption peaks at $\lambda = 300 , 410 ~ nm$. Hence the actual dielectric constant of gold can be expressed as the sum
\begin{equation}
\epsilon_m (\omega) = \epsilon_{intra}(\omega) + \epsilon_{inter}(\omega),
\end{equation}
where $\epsilon_{intra} (\omega)$ is given by Eq. (\ref{Free_Electron_Model_Susceptibility}).

\subsection{Interband transitions}

In what follows, we review the  perturbative theory for the interband transitions of electrons from the $d$-band to the conduction band, describing the effective interband contribution to the dielectric constant of gold $\epsilon_{inter}(\omega)$. This contribution is strongly dependent on the electron temperature $T_{e}$. We start our analysis from the determination of the transition probability of a single electron from the valence to the conduction band, calculating the temperature-dependent interband absorption rate as a function of the optical frequency. In such a derivation, only direct transitions are accounted for and {\em umklapp} processes are neglected \cite{Ashcroft_Book}.

\begin{table}[b]
\centering
\begin{tabular}{ c | c | c | c | c | c | c | }
      & $E_{0v}(eV)$ & $E_{0c}(eV)$ & $m_{v\bot}/m_e$ & $m_{c\bot}/m_e$ & $m_{v\doubleslash}/m_e$ & $m_{c\doubleslash}/m_e$  \\ \hline
  $X$ &   $-1.495$   &   $1.466$    &    $3.500$      &     $0.220$     &    $3.700$    &     $0.120$    \\ \hline
  $L$ &   $-2.380$   &   $-0.390$   &    $0.862$      &     $0.220$     &    $0.804$    &     $0.251$    \\
\end{tabular}
\caption{Band parameters for the $X$ and $L$ transitions used in our calculations ($m_e$ is the electron mass) \cite{Christensen_PRB_1971,Guerrisi_PRB_1975}.}
\label{tab_myfirsttable}
\end{table}

\begin{figure}
\centering
\begin{center}
\includegraphics[width=0.3\textwidth]{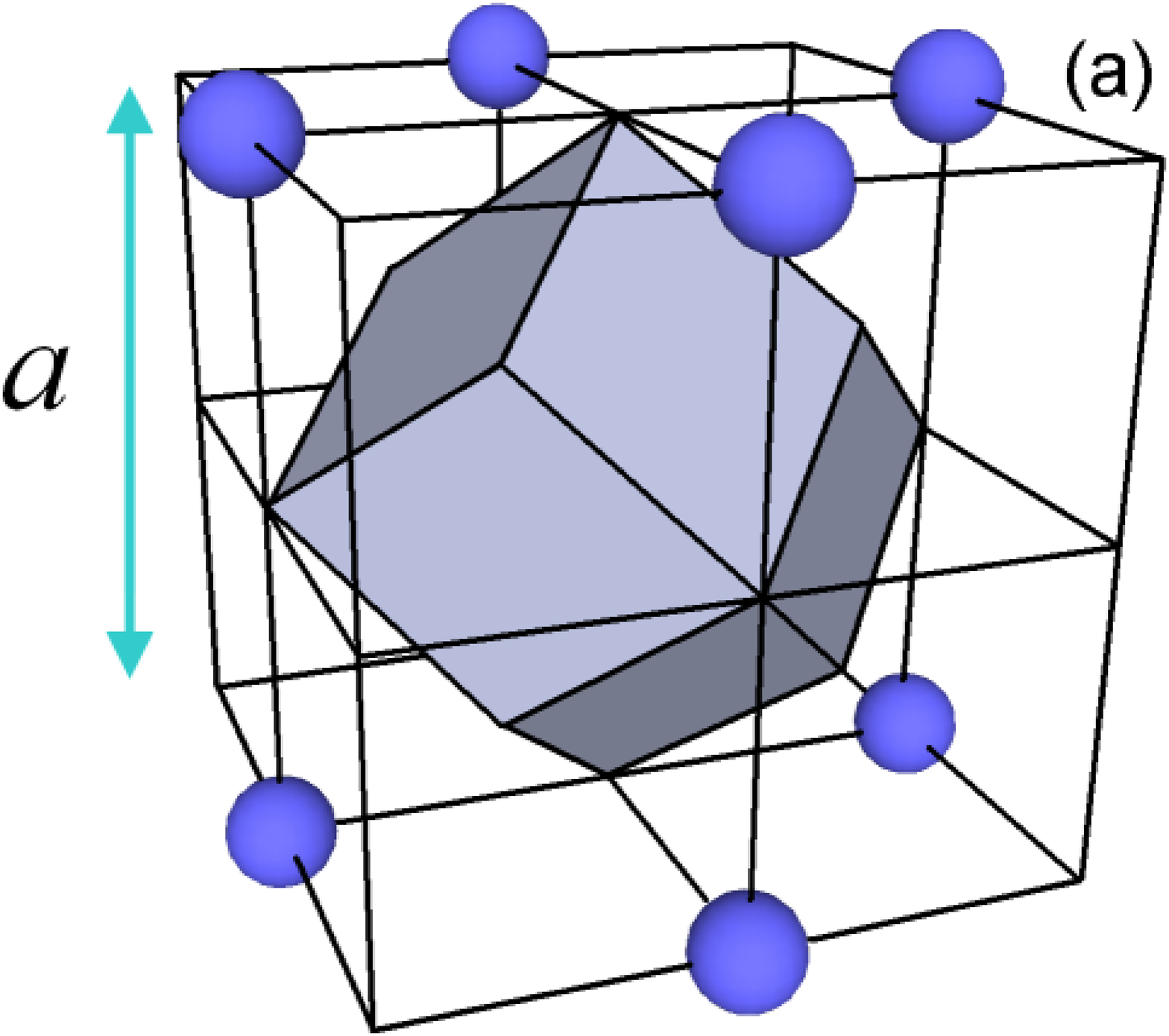}
\includegraphics[width=0.3\textwidth]{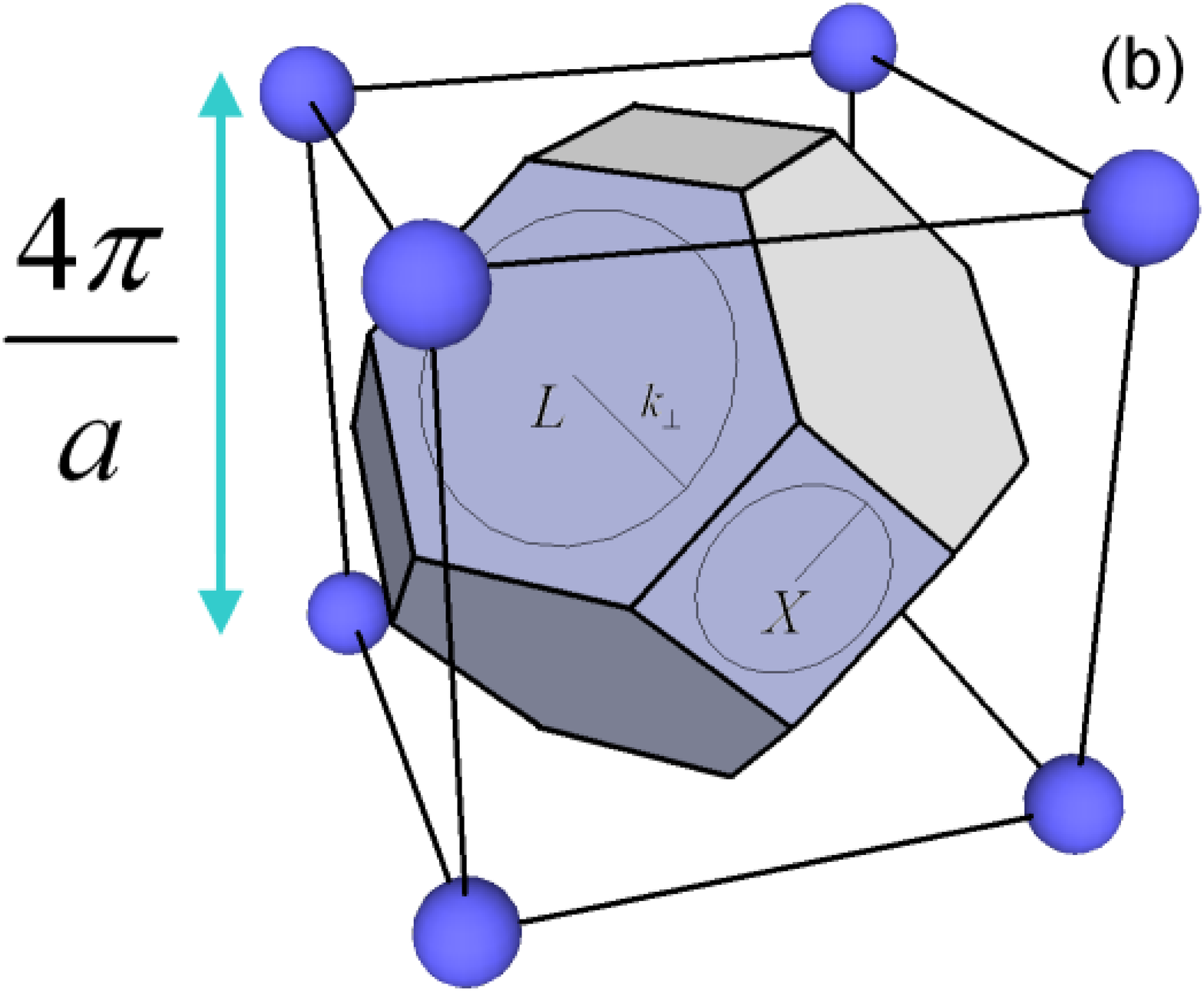}
\caption{(a) Real and (b) reciprocal lattice structures of gold.}
\label{Structure_Fig}
\end{center}
\end{figure}

Gold is characterized by a face centered cubic (f.c.c.) lattice structure, sketched in Fig. \ref{Structure_Fig}a, where the Wigner-Seitz primitive cell is a rhombic dodecahedron and the lattice constant is $a=4.08{\text{\normalfont\AA}}$ \cite{Ashcroft_Book}. The reciprocal lattice, depicted in Fig. \ref{Structure_Fig}b, is body centered cubic (b.c.c.) where the Wigner-Seitz primitive cell is a truncated octahedron. At optical frequencies, the interband absorption is resonant around the points X and L in reciprocal space \cite{Guerrisi_PRB_1975}, which correspond respectively to the centers of the square and hexagonal facets of the truncated octahedron (see Fig. \ref{Structure_Fig}b). Notably, such points are highly symmetric and around them the lattice vector $\vec{k}$ can be expressed as the sum $\vec{k}=\vec{k}_{\bot}+\vec{k}_{\doubleslash}$, where $\vec{k}_{\bot}$ lies on the square $(X)$ or hexagonal $(L)$ facets and $\vec{k}_{\doubleslash}$ is perpendicular to them (see Fig. \ref{Structure_Fig}b). In addition, the Fermi surface has cylindrical symmetry around the $X,L$ points \cite{Ashcroft_Book} and the valence and conduction bands can be approximated in $(k_{\bot},k_{\doubleslash})$ space by elliptic and hyperbolic paraboloids \cite{Guerrisi_PRB_1975}:
\begin{eqnarray}
&& E_v (\vec{k}) = E_{0v} - \frac{\hbar^2k_{\bot}^2}{2m_{v\bot}} - \frac{\hbar^2k_{\doubleslash}^2}{2m_{v\doubleslash}}, \label{EVKBNDFRM}\\
&& E_c (\vec{k}) = E_{0c} + \frac{\hbar^2k_{\bot}^2}{2m_{c\bot}} - \frac{\hbar^2k_{\doubleslash}^2}{2m_{c\doubleslash}}, \label{ECKBNDFRM}
\end{eqnarray}

where $v,c$ indicate the valence and conduction bands. The values of the constants $E_{0v}$, $E_{0c}$, $m_{v\bot}$, $m_{c\bot}$, $m_{v\doubleslash}$, $m_{c\doubleslash}$ for the $X$ and $L$ transitions that we use in our calculations are listed in Table \ref{tab_myfirsttable}. Note that, as a consequence of the cylindrical symmetry of the Fermi surface, the conduction and valence bands around the $X,L$ points do not depend on the direction of $\vec{k}_{\bot}$, but only on its modulus $k_{\bot}$. The Fermi level $E_F$ lies in the conduction band $E_c (\vec{k})$ and for the sake of simplicity (and without loss of generality) we perform a constant shift of all energies and assume that $E_F=0$.

\begin{figure}
\centering
\begin{center}
\includegraphics[width=0.45\textwidth]{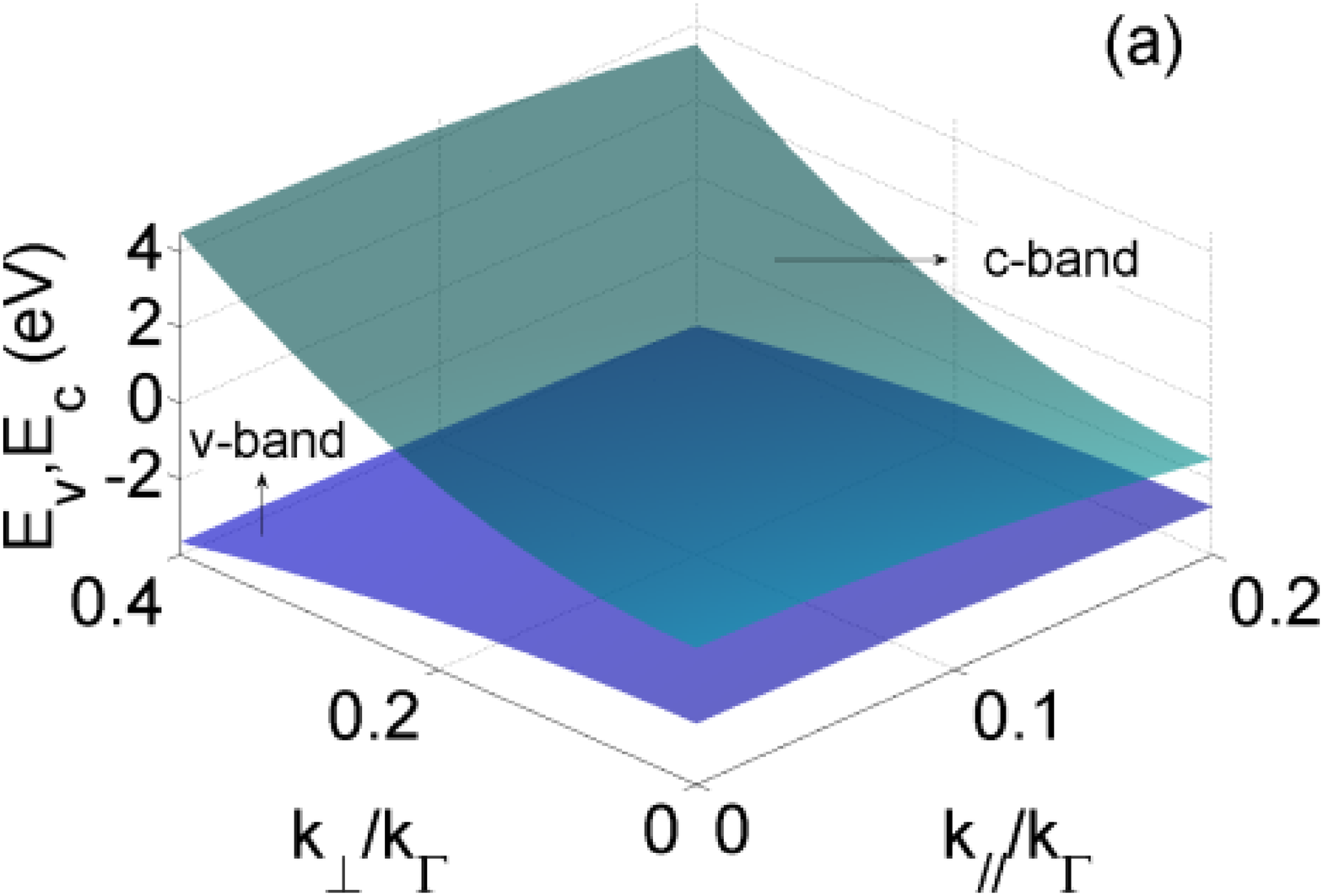}
\includegraphics[width=0.45\textwidth]{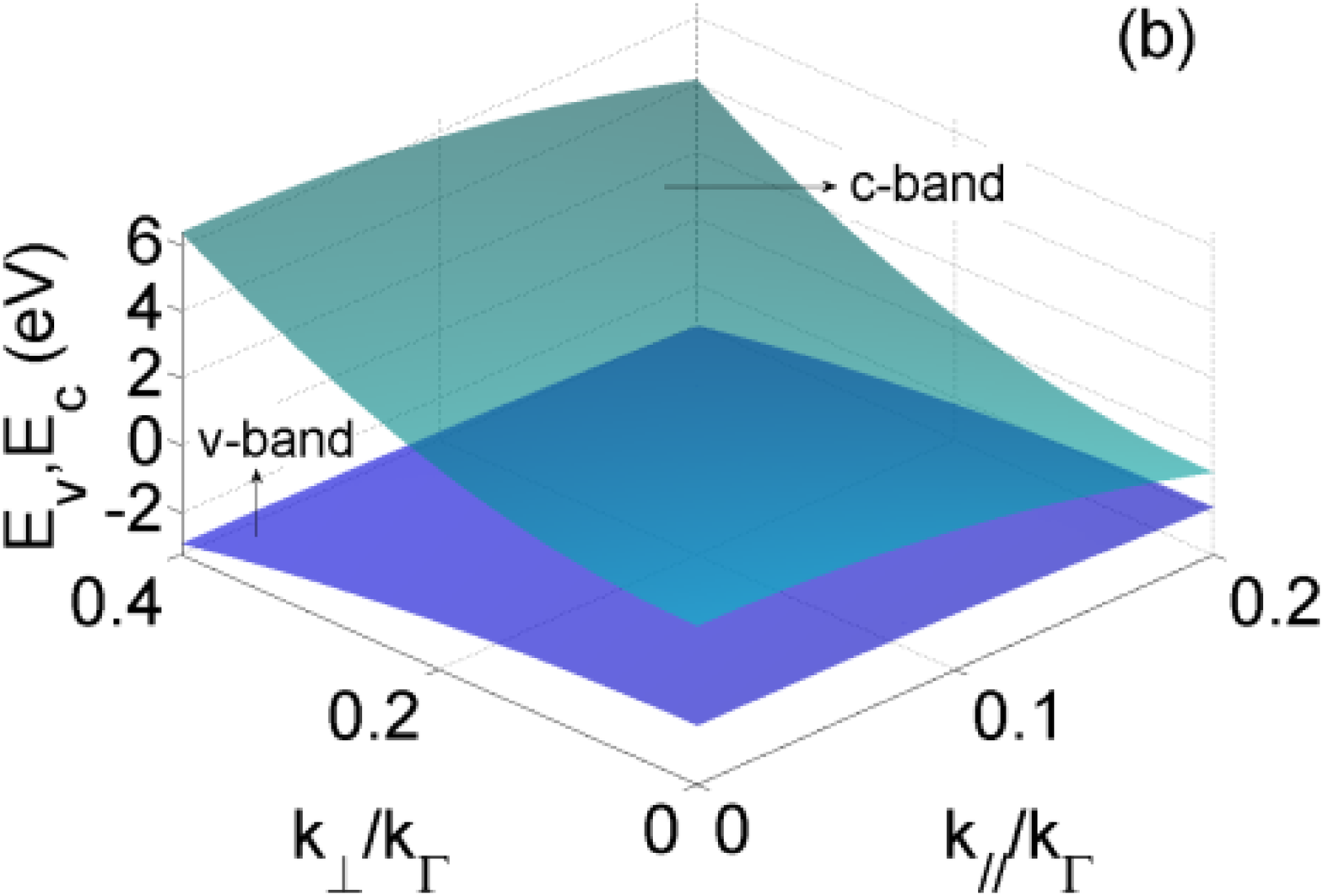}
\caption{Valence and conduction bands around the (a) $L$ and (b) $X$ points as functions of the lattice moduli $k_{\bot},k_{\doubleslash}$. The upper surfaces represent the conduction bands and the lower surfaces the valence bands. The quantities $k_{\bot},k_{\doubleslash}$ are normalized to $k_{\Gamma}=\sqrt{3}\pi/a$, the distance between the $X,L$ points and the center of the Brillouin zone $(\Gamma)$.}
\label{X_L_Band_Structure_Fig}
\end{center}
\end{figure}

The valence and conduction bands $E_v(k_{\bot},k_{\doubleslash})$, $E_c(k_{\bot},k_{\doubleslash})$ are plotted as functions of the moduli $k_{\bot},k_{\doubleslash}$ in Figs. \ref{X_L_Band_Structure_Fig}a,b for the $X$ and $L$ points. In this figure, the quantities $k_{\bot},k_{\doubleslash}$ are measured in terms of $k_{\Gamma}=\sqrt{3}\pi/a$, which is the distance between the $X,L$ points and the center of the Brillouin zone $(\Gamma)$ \cite{Ashcroft_Book}. The upper surfaces correspond to the conduction bands while the lower surfaces correspond to the valence bands. Note that the paraboloid approximation made in Eqs. (\ref{EVKBNDFRM},\ref{ECKBNDFRM}) is accurate only if $k_{\bot},k_{\doubleslash}\ll k_{\Gamma}$. Note also that every point in the fictitious $(k_{\bot},k_{\doubleslash})$ space corresponds to a circle of radius $k_{\bot}$ at a distance $k_{\doubleslash}$ from either the $X$ or the $L$ points in reciprocal $\vec{k}$-space. The quantum states of electrons in the valence and conduction bands are Bloch wavefunctions
\begin{eqnarray}
\psi_{v,c} = \Omega^{-1/2} u_{\vec{k},v,c}(\vec{r})\exp(i\vec{k}\cdot\vec{r}),
\end{eqnarray}
where $\Omega$ is the primitive cell volume and $\vec{k}$ is the Brillouin wavevector. Fermi's golden rule provides the probability per unit time of transitions from valence to conduction band \cite{Cohen_Book}:
\begin{equation}
R_{v,c} (\vec{k}) = \frac{\pi e^2|E_0|^2|\vec{p}_{v,c}|^2}{6\hbar m_e^2\omega^2}\delta \left[ E_c(\vec{k}) - E_l(\vec{k}) -\hbar\omega \right], \label{FrmGldnRlEq}
\end{equation}
where $E_0$ is the electric field amplitude and
\begin{equation}
\vec{p}_{v,c} = - i \frac{\hbar}{\Omega} \int_{\Omega} d^3 r \left[ u^*_{\vec{k},v}(\vec{r}) \nabla u_{\vec{k},c}(\vec{r}) \right] .
\end{equation}
In Eq. (\ref{FrmGldnRlEq}) {\em umklapp} processes have been neglected and the Dirac delta-function ensures conservation of energy for the direct transitions ($\vec{k}$ is conserved). To obtain the transition rate from the initial to the final band one needs to sum over all available states, labeled by the $\vec{k}$ vector. The sum over $\vec{k}$ can be replaced by integration if the density of states $D(\vec{k})=2\Omega/(2\pi)^3$ is introduced. Thus, the transition rate per unit volume is
\begin{equation}
W_{c,v} = \frac{2}{(2\pi)^3} \int_{BZ}d^3k R_{c,v}f(E_v,T_e)[1-f(E_c,T_e)],
\end{equation}
where $T_e$ is the electronic temperature, $f(E,T_e)$ is the Fermi-Dirac occupation number and the integration is taken over the volume of the first Brillouin zone. Note that the Pauli exclusion principle for every $v \rightarrow c$ direct transition is carefully considered in the expression for $W_{c,v}$. Indeed, the factor $f[E_v(\vec{k}),T_e]$ accounts for the probability that the $\vec{k}$-state in the valence band is occupied, while the factor $1-f[E_c(\vec{k}),T_e]$ accounts for the probability that the $\vec{k}$-state in the conduction band is empty. The absorbed power per unit volume $P_A$ can be directly related to the imaginary part of the dielectric constant \cite{Landau_Book}:
\begin{equation}
P_A = W_{c,v}\hbar\omega = \frac{1}{2}\epsilon_0\omega\epsilon''_{inter}|E_0|^2.
\end{equation}

Hence, the imaginary part of the dielectric constant due to the interband transitions is explicitly given by
\begin{equation}
\epsilon''_{inter}(\omega,T_e) = \frac{\pi e^2 |\vec{p}_{c,v}|^2}{3\epsilon_0 m_e^2 \omega^2} J_{c,v}(\omega,T_e) \label{realparteps},
\end{equation}
where we have approximated the matrix element $\vec{p}_{c,v}$ to be independent of $\vec{k}$, which is true only in the limit $|\vec{k}|\ll k_{\Gamma}$.
$J_{c,v}(\omega,T_e)$ provides the number of available direct $v \rightarrow c$ transitions responsible for interband absorption. For this reason, this quantity is usually named the joint density of states (JDOS):
\begin{eqnarray}
J_{c,v}(\omega,T_e) & = & \frac{2}{(2\pi)^3} \int_{BZ} d^3 k \left\{\delta\left[ E_c(\vec{k}) - E_v(\vec{k})  - \hbar \omega \right]\times\right. \nonumber \\
                       && \left. \times f [ E_v(\vec{k}) , T_e ] ( 1 - f [ E_c(\vec{k}) , T_e ] ) \right\} \label{JDOS_ANAL_EXPR_GEN}.
\end{eqnarray}
The real part of the dielectric constant can then be obtained directly from Eq. (\ref{realparteps}) using the Kramers-Kronig relation
\begin{equation}
\epsilon'_{inter}(\omega,T_e) = \frac{1}{\pi} {\cal P} \int_{-\infty}^{+\infty} \frac{\epsilon''_{inter} (\omega',T_e)}{\omega'-\omega} d \omega',\label{KKR_EQ}
\end{equation}
where ${\cal P}$ represents the principal value of the integral.

\begin{figure}
\centering
\begin{center}
\includegraphics[width=0.45\textwidth]{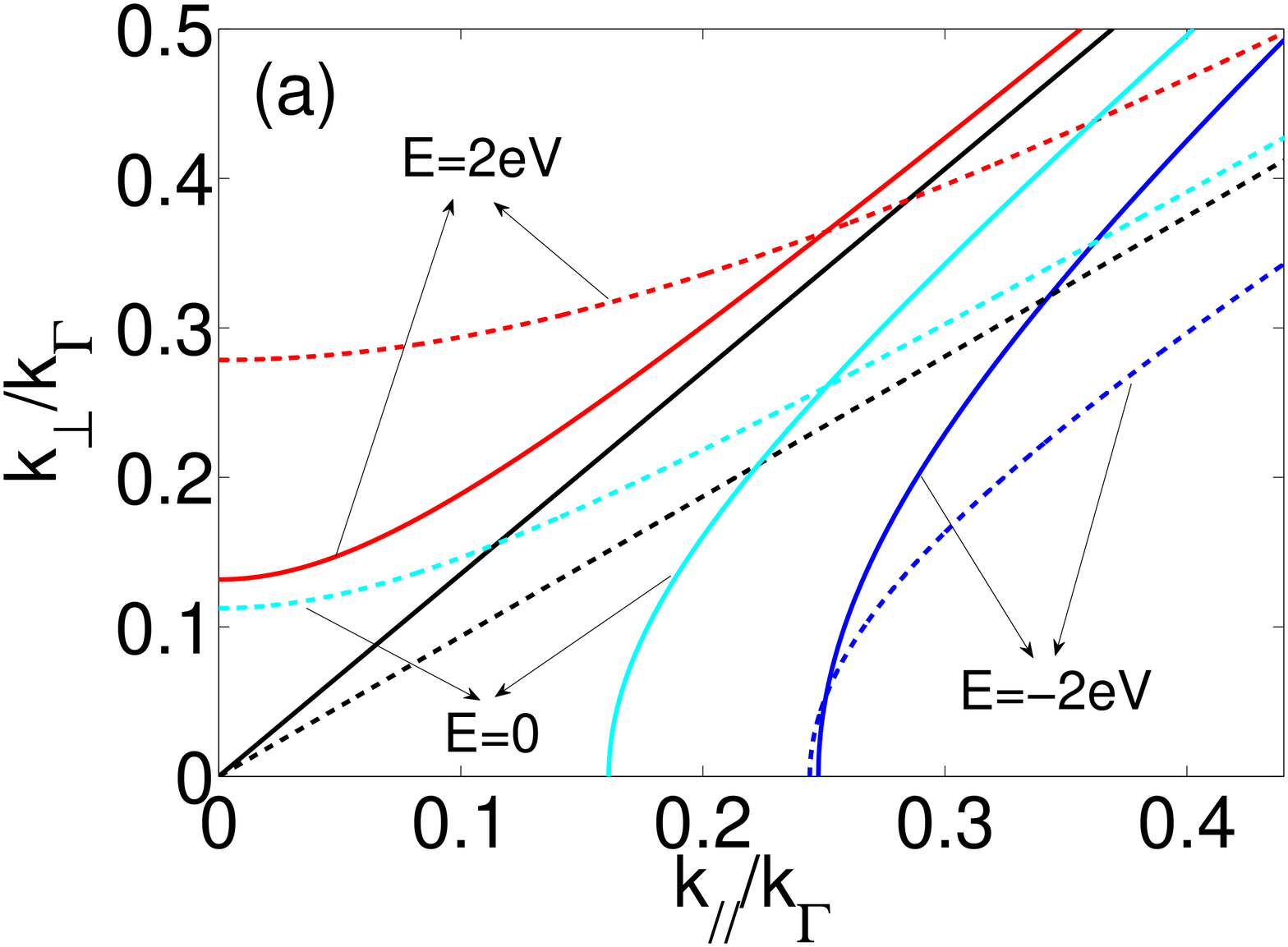}
\includegraphics[width=0.45\textwidth]{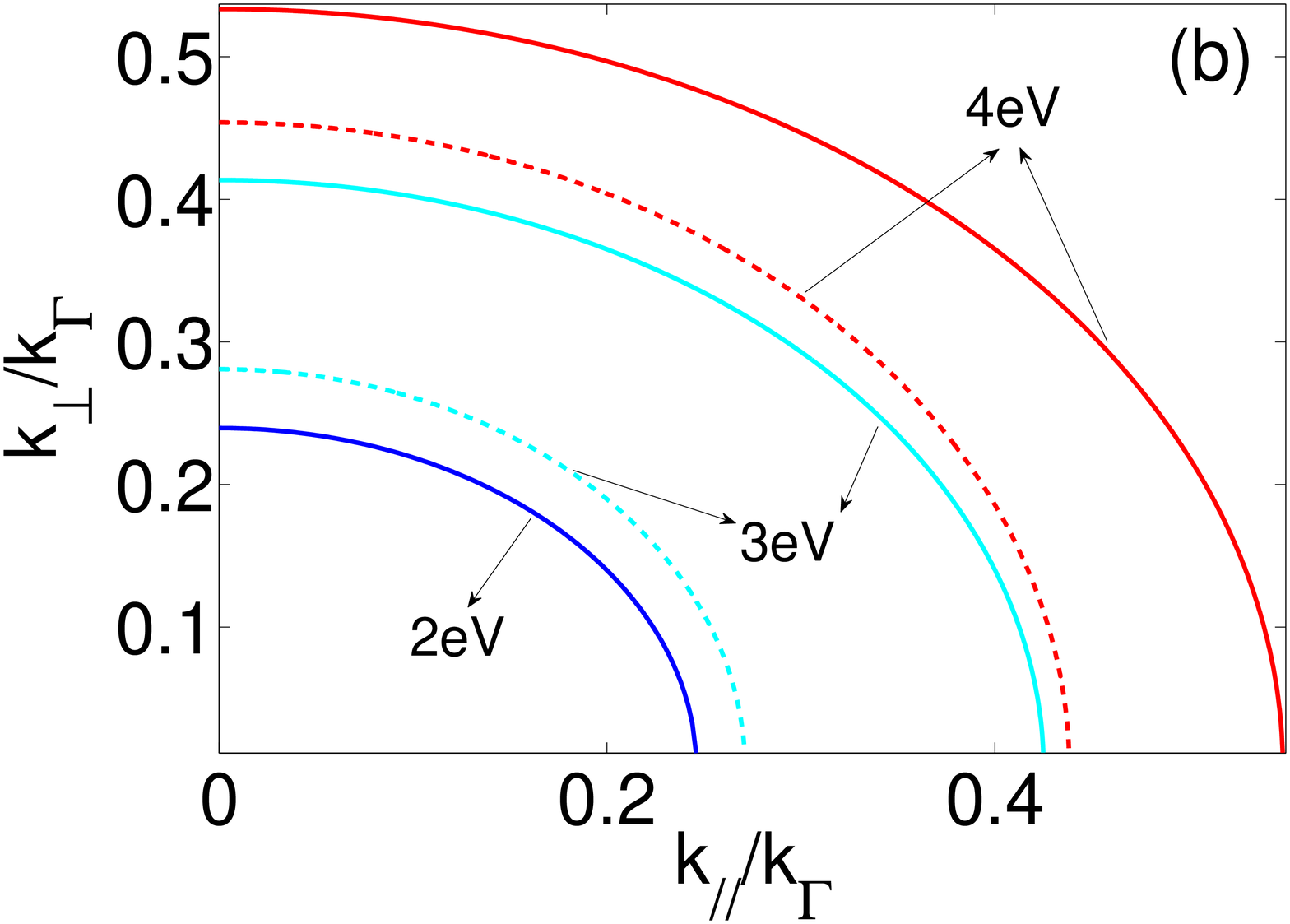}
\caption{(a) Contour-plots of the CECBS hyperbolae in the $(k_{\bot},k_{\doubleslash})$ plane. Blue, cyan and red curves correspond to $E=-2,0,+2eV$, while full and dashed curves refer to the $X$ and $L$ transitions. The full and dashed black lines correspond to the asymptotes of the $X,L$ hyperbolae. (b) Contour-plots of the CEDS ellipses in the $(k_{\bot},k_{\doubleslash})$ plane. Blue, cyan and red curves correspond to $\hbar\omega-E=2,3,4 ~ eV$ and full and dashed curves refer to the $X$ and $L$ transitions.}
\label{CEDS_Fig}
\end{center}
\end{figure}

\subsection{Calculation of the JDOS and of its thermo-derivative}

An exact analytical calculation of the JDOS, given by Eq. (\ref{JDOS_ANAL_EXPR_GEN}), is not possible to the best of our knowledge, so in this section we calculate the JDOS and its thermo-derivative numerically. In turn, we use the JDOSs of the $X,L$ transitions to calculate the total interband contribution to the dielectric constant and its dependence on the electronic temperature $T_e$.

The calculation of the JDOS can be greatly simplified by introducing another quantity, the energy distribution of the JDOS (EDJDOS) \cite{Chrinstensen_PSSB_1972}, defined as
\begin{eqnarray}
D(E,\omega) & \equiv & \frac{2}{(2\pi)^3} \int d^3 k \delta\left( E_c - E_v -\hbar \omega \right) \delta\left(E_c - E \right) = \nonumber \\
      & = & \frac{2}{(2\pi)^3} \oint_{\cal A} \frac{dl}{\left| \nabla_{\vec{k}} E_v \times \nabla_{\vec{k}} E_c \right|}, \label{EDJDOS_Circ_Int}
\end{eqnarray}
where the line-integral is taken over the closed path ${\cal A}$, given by the intersection of the constant energy of the conduction band surface (CECBS): $E_c (\vec{k}) = E$, with the the constant energy difference surface (CEDS): $\hbar\omega - E + E_v (\vec{k}) = 0$. The JDOS can be expressed in terms of the EDJDOS as the integral
\begin{eqnarray}
J_{c,v}(\omega,T) & = &  \int_{E_{min}}^{E_{max}} dE \left\{ D(E,\omega)\times \right. \label{JDOS_INT_EXPR}\\
                       &   &  \left. \times f(E-\hbar\omega,T_e)\left[1-f(E,T_e)\right] \right\}. \nonumber
\end{eqnarray}
At every constant value of conduction energy $E$, the CECBS is a hyperbola in the plane of the moduli $(k_{\bot},k_{\doubleslash})$ and correspondingly a hyperboloid in $\vec{k}$-space. Some contour-plots of the CEBCS hyperbolae for the $X$ (full curves) and $L$ (dashed curves) transitions are depicted in Fig. \ref{CEDS_Fig}a. Blue, cyan and red curves correspond to the energy values $E=-2,0,+2eV$, indicated with arrows. The full and dashed black lines represent the asymptotes of the $X,L$ hyperbolae. Note that the cyan curve represents the Fermi level and that the concavity of the hyperbolae depends on the sign of $E-E_{0c}$. For every fixed $E,\omega$ values such that $E-\hbar\omega \le E_{0v}$, the CEDS is an ellipse in the plane of the moduli $(k_{\bot},k_{\doubleslash})$ and correspondingly an ellipsoidal cap in $\vec{k}$-space. Some contour-plots of the CEDS ellipses for the $X$ (full curves) and $L$ (dashed curves) transitions are depicted in Fig. \ref{CEDS_Fig}b. Blue, cyan and red curves correspond to the energy differences $\hbar\omega-E=2,3,4eV$, also indicated with arrows. The blue dashed line is absent since the condition for existence of the $L$ ellipse is not fulfilled at $\hbar\omega-E = 2eV$.

\begin{figure}
\centering
\begin{center}
\includegraphics[width=0.45\textwidth]{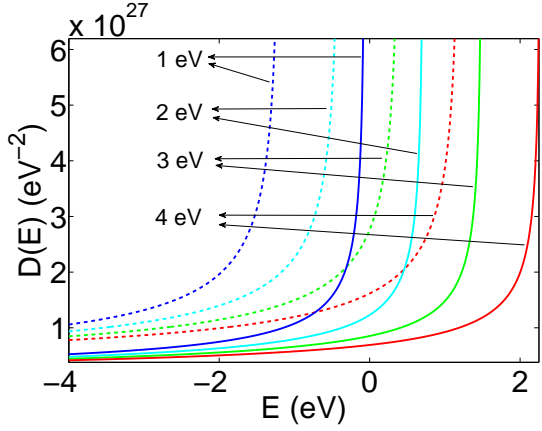}
\caption{EDJDOS of gold as a function of energy $E$ in the conduction band for $X$ (full curves) and $L$ (dashed curves) interband transitions. Blue, cyan, green and red curves correspond to $\hbar\omega=1,2,3,4 ~ eV$.}
\label{EDJDOS_Fig}
\end{center}
\end{figure}

In Eq. (\ref{EDJDOS_Circ_Int}), the volume integral in the reciprocal space has been reduced to a circuit integral over the closed integration path ${\cal A}$ as a consequence of the Dirac delta-functions $\delta\left( E_c - E_v -\hbar \omega \right), \delta\left( E_c - E \right)$. Hence, the integration path ${\cal A}$ is a circle of radius $k_{\bot}$ displaced at a distance $k_{\doubleslash}$ from the $X,L$ points in the reciprocal space, resulting from the intersection of the CECBS (an hyperboloid) and the CEDS (an ellipsoidal cap). Such a circle in $\vec{k}$-space corresponds to the point $(k_{\bot},k_{\doubleslash})$ in the space of the moduli. The circuit integral in Eq. (\ref{EDJDOS_Circ_Int}) can be solved straightforwardly, leading to the following expression for the EDJDOS:
\begin{equation}
D (E,\omega) = \frac{g{\cal E}^{-3/2}}{8\pi^2\sqrt{\eta}}\frac{\theta\left[{\cal V}(\omega) - E\right]}{\sqrt{{\cal V}(\omega) - E}},
\end{equation}
where $g$ is a degeneracy number related to the number of X,L points in the first Brillouin zone ($g=6,8$ for $X,L$ transitions), $E_G = E_{0c} - E_{0v}$, $\theta(x)$ is the Heaviside step function and
\begin{eqnarray}
\eta             & = & 1 + m_{c\bot}/m_{v\bot}, \nonumber \\
{\cal V}(\omega) & = & \eta^{-1}(\hbar\omega + \eta E_{0c} - E_G), \\
{\cal E}^{-3/2}  & = & \frac{2m_{c\bot}}{\hbar^3} \sqrt{ \frac{ 2 m_{v\doubleslash} m_{v\bot} m_{c\doubleslash} }{ m_{v\doubleslash}m_{c\bot} + m_{v\bot}m_{c\doubleslash}} }. \nonumber
\end{eqnarray}
Note that ${\cal E}$ has the physical dimension of energy $(eV)$, so that $D(E,\omega)$ has the physical dimension of the inverse of energy squared $(eV^{-2})$. The EDJDOS $D(E)$ of gold is plotted as a function of the energy level of the conduction band $E$ in Fig. \ref{EDJDOS_Fig} for different values of $\hbar\omega$. Note that the EDJDOS is singular at the points $E={\cal V}(\omega)$, known in solid state physics as Van Hove singularities \cite{Ashcroft_Book}.

As can be understood from Figs. \ref{CEDS_Fig}a,b, the neck of the CECBS hyperbola grows with energy $E$ while for every fixed frequency $\omega$ the vertical semi-axis of the CEDS ellipse shrinks as $E$ increases. Thus, there exists a maximum energy $E_{max}$ where the CEDS ellipse is either tangential to the CECBS hyperbola on its neck or the semi-axes of the CEDS ellipse vanish:
\begin{equation}
E_{max}(\omega) = (\sigma - \kappa\hbar\omega)\theta(E_G-\hbar\omega) + {\cal V}(\omega)\theta(\hbar\omega-E_G),
\end{equation}
where $\kappa = m_{v\doubleslash}/(m_{c\doubleslash}-m_{v\doubleslash})$ and $\sigma = E_{0c} + \kappa E_G$. Conversely, as a consequence of the paraboloid approximation for the conduction band, the lower integration boundary $E_{min}(\omega)$ remains arbitrary. We set the minimum energy to
\begin{equation}
E_{min}(\omega) = {\cal V}(\omega)-\frac{\hbar^2k_l^2}{2{\cal M}},
\end{equation}
where $k_l=k_{\Gamma}/5$ and
\begin{equation}
{\cal M} = \frac{m_{v\doubleslash}m_{c\doubleslash}(m_{v\bot}+m_{c\bot})}{m_{v\bot}m_{c\doubleslash}+m_{v\doubleslash}m_{c\bot}}.
\end{equation}
The choice of the lower integration extremum is critical. Indeed, by choosing a small value for $k_l$ one neglects the dispersion of the valence band; on the other hand, by choosing a large value of $k_l$, the constant matrix element and parabolic band approximations cease to be valid.

The problem of calculating the JDOS for the $X,L$ interband transitions of gold is then reduced to the calculation of the integral in Eq. (\ref{JDOS_INT_EXPR}), which unfortunately has no simple analytical solution.  By labeling the JDOSs of the $X,L$ transitions with $J_{c,v}^{X},J_{c,v}^{L}$, the total interband correction to the imaginary part of the dielectric constant of gold is given by the sum
\begin{equation}
\epsilon''_{inter}(\omega,T_e) = \frac{\pi e^2 }{3\epsilon_0 m_e^2 \omega^2} \left[ |\vec{p}_{c,v}^{X}|^2 J_{c,v}^{X} + |\vec{p}_{c,v}^{L}|^2 J_{c,v}^{L} \right].
\end{equation}

\begin{figure}
\centering
\begin{center}
\includegraphics[width=0.45\textwidth]{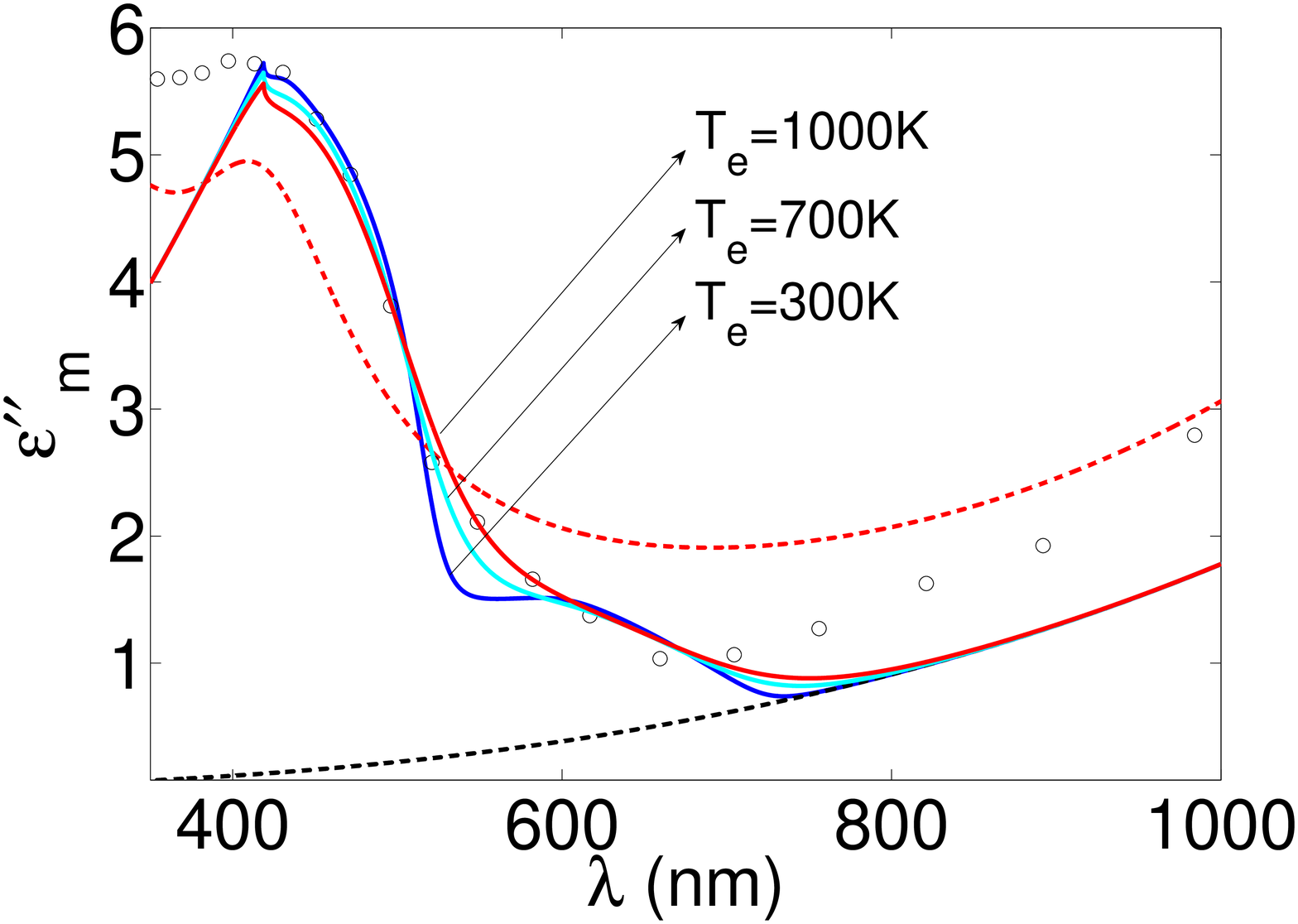}
\caption{Results of a numerical calculation of the imaginary part of the dielectric constant of gold $\epsilon''_m$ as a function of the wavelength $\lambda$. The open circles are the experimental data points of Johnson and Christy \cite{Johnson1972}. The red dashed curve corresponds to a fit to the experimental data in Ref. \cite{Ung2007}, while the black dashed curve is the intraband contribution given by Eq. (\ref{Free_Electron_Model_Susceptibility}). The full blue, cyan and red curves correspond to the sum of the interband and intraband contributions $\epsilon''_m=\epsilon''_{intra}+\epsilon''_{inter}$ for electronic temperatures $T_e=300,700,1000$ $^{o}$K. }
\label{Eps2_Interband}
\end{center}
\end{figure}

The real part of the interband dielectric susceptibility $\epsilon'_{inter}(\omega,T_e)$ can be calculated from Eq. (\ref{KKR_EQ}). We have computed $J_{c,v}^{X},J_{c,v}^{L}$ and $\epsilon''_{inter}(\omega,T_e)$ numerically. The results of these calculations are plotted in Fig. \ref{Eps2_Interband}. For the dipole matrix elements $|\vec{p}_{c,v}^{X}|^2,|\vec{p}_{c,v}^{L}|^2$ we have used the values in Refs. \cite{Guerrisi_PRB_1975,Suffczynski_1960} ($ g_L |\vec{p}_{c,v}^{L}|^2 = 1.6015 \times 10^{-47} J \times kg  , g_X |\vec{p}_{c,v}^{X}|^2 = 0.321 \times g_L |\vec{p}_{c,v}^{L}|^2$, where $g_L=8,g_X=6$). In Fig. \ref{Eps2_Interband}, the numerical calculation of $\epsilon''_m=\epsilon''_{intra}+\epsilon''_{inter}$ is plotted at several electronic temperatures $T_e=300,700,1000$ $^{o}$K (blue, cyan and red full curves). The open circles represent the experimental data of Johnson and Christy \cite{Johnson1972}, the red dashed line is a fit to the experimental data in Ref. \cite{Ung2007} and the black dashed line represents the intraband contribution $\epsilon_{intra}$. Note that the numerical results fit quite well to the experimental data of Johnson and Christy for $400 nm < \lambda < 1 \mu m $. In particular, a very good fit to the measurements is obtained for $T_e=700$ $^{o}$K. For $\lambda < 410 nm$, other interband transitions become important and the contribution of the $X,L$ points is not sufficient to explain the experimental measurements.

\begin{figure}
\centering
\begin{center}
\includegraphics[width=0.45\textwidth]{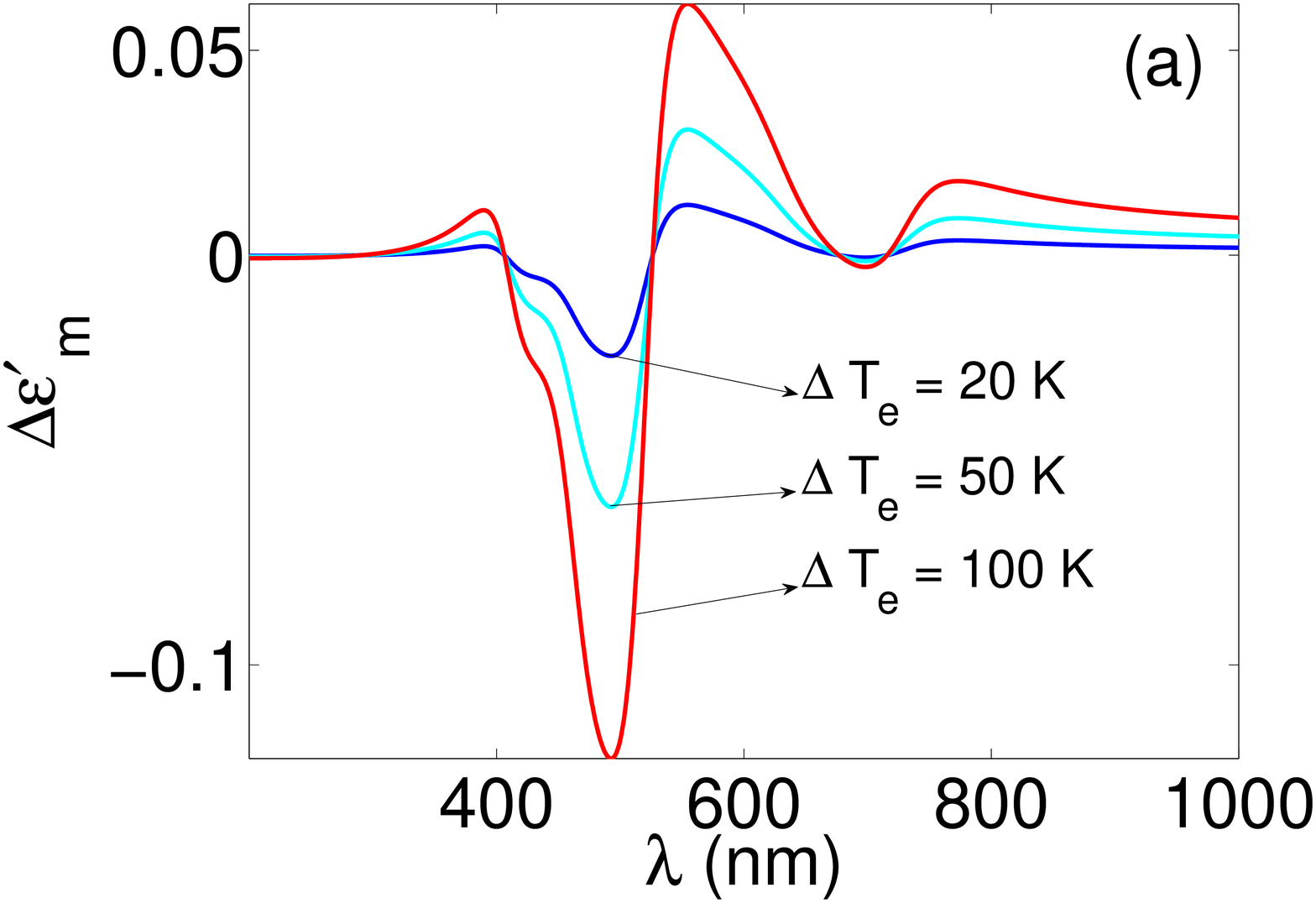}
\includegraphics[width=0.45\textwidth]{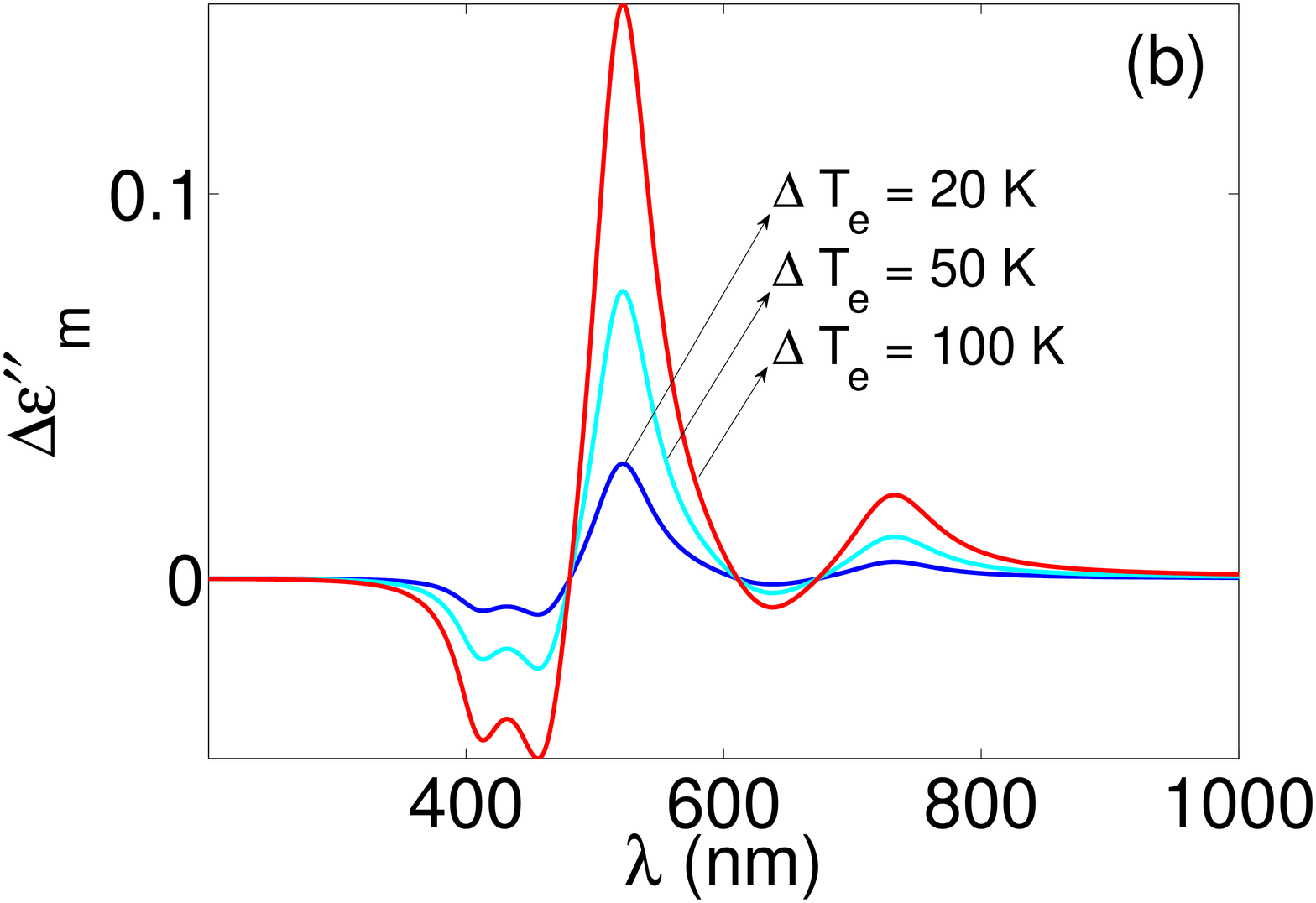}
\caption{Thermo-modulation of the dielectric constant of gold. (a) Real and (b) imaginary corrections to the dielectric constant $\Delta\epsilon_m=\partial_{T_e}\epsilon_{inter}(\omega,T_0)(T_e-T_0)$ ($T_0=300$ $^{o}$K). Blue, cyan and red curves correspond to the electronic temperatures $T_e = 320, 350, 400$ $^{o}$K.}
\label{Delteps_Fig}
\end{center}
\end{figure}

If the conduction electrons are taken out of equilibrium by light-induced heating, the interband absorption is affected by the so called {\it Fermi smearing effect} \cite{Guerrisi_PRB_1975}. Increasing temperature broadens the electron distribution around the Fermi energy, modifying the effective optical properties of the metal. In order to understand the temperature dependence of $\epsilon''_{inter}$, contained in the $X,L$ JDOSs, one needs to compute numerically the thermo-derivatives $\partial_{T_e}J_{c,v}^{X,L}(\omega,T_e)$. The resulting interband dielectric thermo-derivative can be fitted to a series of five Lorentzian functions:
\begin{equation}
\partial_{T_e}\epsilon''_{inter}(\omega) = \sum_{j=1}^5\frac{{\cal F}_{j}\omega_p^2}{(\omega-\omega_{j})^2+\gamma_{j}^2},\label{ThermDerivFitEq}
\end{equation}
where $\omega_p=1.1515\times 10^{16} rad/sec$ is the plasma frequency of gold, calculated by fitting to the experimental data in the far-infrared \cite{Ung2007}. The fit parameters are given in Table \ref{tab_mysecondtable}. If a simplified Lorentzian behaviour is assumed, the calculation of the real part $\partial_{T_e}\epsilon'_{inter}$ through the integration of the Kramers-Kronig relation given by Eq. (\ref{KKR_EQ}) is straightforward:
\begin{equation}
\partial_{T_e}\epsilon'_{inter}(\omega) = \sum_{j=1}^5\frac{{\cal F}_{j} \omega_p^2 (\omega_{j} - \omega)} {\gamma_{j}(\omega-\omega_{j})^2+\gamma_{j}^3}.
\end{equation}
The exact calculation of the complex dielectric thermo-derivative $\partial_{T_e}\epsilon_{inter}(\omega)$ is of fundamental importance for the description of light-induced self-thermo-modulation. Basically, as light impinges on the gold surface, the electrons in the conduction band are heated and the dielectric constant is modified by the amount
\begin{equation}
\Delta\epsilon_m(\omega) = \partial_{T_e}\epsilon_{inter}(\omega)\Delta T_e. \label{ThermoModDelteps}
\end{equation}
Note that this thermo-modulational process is intrinsically nonlinear, since the increase of temperature depends on the absorbed optical power  $\Delta T_e (P_A)$. The spectral dependence of the complex correction $\Delta\epsilon_m(\omega)$ is plotted in Figs. \ref{Delteps_Fig}a,b for several values of temperature variation $\Delta T_e = 20, 50 , 100$ $^{o}$K (blue, cyan and red curves respectively). Note that the spectral dependence of both real and imaginary parts $\Delta\epsilon'_m(\omega),\Delta\epsilon''_m(\omega)$ is non-trivial and they can be either positive or negative; hence the optical absorption can increase or decrease, depending on the wavelength $\lambda$.

\begin{table}[t]
\centering
\begin{tabular}{ c | c | c | c | c| c | }
           $j$             &    $1$    &    $2$    &    $3$    &     $4$   &     $5$   \\ \hline
${\cal F}_{j}\times 10^7$  & $-1.6969$ & $-2.9413$ & $+5.0681$ & $-1.0016$ & $+0.4045$ \\ \hline
$\omega_{j}/\omega_p$      &  $0.3982$ & $0.3541$  &  $0.3140$ &  $0.2587$ & $0.2238$  \\ \hline
$\gamma_{j}/\omega_p $     &  $0.0217$ & $0.0216$  &  $0.0173$ &  $0.0217$ & $0.0130$  \\ \hline
\end{tabular}
\caption{Fit parameters for the thermo-derivative $\partial_{T_e}\epsilon''_{inter}(\omega)$, given by Eq. (\ref{ThermDerivFitEq}). $\omega_p=1.1515\times 10^{16} rad/sec$ is the plasma frequency of gold, which has been calculated by fitting with the free-electron model  the experimental data in the far-infrared \cite{Ung2007}.}
\label{tab_mysecondtable}
\end{table}

\section{Electron temporal dynamics and the two-temperature model}

As we have shown in the previous section, an optical beam impinging on a metal surface modifies the effective interband susceptibility by heating the electrons in the conduction band. This light-induced electron heating can be described through the two temperature model (TTM) \cite{Anisimov_JETP_1975}, which takes account of the energy balance between the conduction electrons and the lattice. The electrons have a relatively small heat capacity and so thermalize through electron-electron collisions with a characteristic time of order $\tau_{th} \approx 300 fs$. If one wishes to describe the temporal electron dynamics for ultrashort optical pulses ($\tau_0 \approx 100 fs$), it is necessary also to include the energy contribution of the non-thermalized electrons in the energy balance. This can be calculated directly from the Boltzmann equation in the relaxation time approximation \cite{Carpene_PRB_2006,Sun_PRB_1994}. A phenomenological description of the electron temporal dynamics can be obtained by separating the electron distribution of energy into thermalized and non-thermalized parts \cite{Fann_PRB_1992,Sun_PRB_1993}:
\begin{eqnarray}
\partial_t N(t)       & = & - ( \gamma_e + \gamma_l ) N(t) + P_A(t), \nonumber \\
C_e \partial_t T_e(t) & = &  {\cal C} ( T_l - T_e ) + \gamma_e N(t), \\
C_l \partial_t T_l(t) & = &  {\cal C} ( T_e - T_l ) + \gamma_l N(t), \nonumber
\end{eqnarray}

\begin{table}[b]
\centering
\begin{tabular}{ c | c | c | c | }
Parameter  &       Value       &      Units      &        Ref.          \\ \hline
  $C_e$    & $2.1 \times 10^4$ & $Jm^{-3}K^{-1}$ & \cite{Lin_SPIE_2006} \\ \hline
  $C_l$    & $2.5 \times 10^6$ & $Jm^{-3}K^{-1}$ & \cite{Lin_SPIE_2006} \\ \hline
${\cal C}$ & $2\times 10^{16}$ &    $s^{-1}$     & \cite{Lin_SPIE_2006} \\ \hline
$\gamma_e$ & $2\times 10^{12}$ &    $s^{-1}$     & \cite{Sun_PRB_1994}  \\ \hline
$\gamma_l$ & $1\times 10^{12}$ &    $s^{-1}$     & \cite{Sun_PRB_1994}  \\ \hline
\end{tabular}
\caption{Parameters of the two temperature model used in our numerical calculations and corresponding references. The electronic and lattice heat capacities are calculated for $T_e = T_l = T_{eq} = 300$ $^{o}$K.}
\label{tab_mythirdtable}
\end{table}

where $P_A(t)$ is the mean absorbed power per unit volume, $T_e(t),T_l(t)$ are the electronic and lattice temperatures, $C_e,C_l$ are the electronic and lattice heat capacities per unit volume and $N(t)$ is the energy density stored in the non-thermalized part of the electronic distribution. When an ultrashort optical pulse impinges on the metal, it is absorbed and transfers energy to the non-thermalized electrons. In turn, the non-thermalized electrons release energy density $\gamma_e N(t)$ to the thermalized electrons via electron-electron scattering and energy density $\gamma_l N(t)$ to the lattice via electron-phonon scattering. The non-thermalized electrons achieve thermal equilibrium with a characteristic time delay of $\tau_{th} = ( \gamma_e + \gamma_l )^{-1}$, where $\gamma_e,\gamma_l$ are the electron and lattice thermalization rates. Once heated by an ultrashort optical pulse, the thermalized electrons gradually release energy to the lattice via electron-phonon scattering, which is accounted for by the coupling coefficient ${\cal C}$. Ultimately, for long times, the electrons reach thermal equilibrium with the lattice. The parameters used in the numerical calculation are given in Table \ref{tab_mythirdtable}. Note that since the lattice heat capacity is much larger than the electron heat capacity, while the temporal variation of the electronic temperature $T_e(t)$ is significant, the lattice temperature $T_l(t)$ does not change significantly with time, i.e., $T_l(t) \approx const$.

Note also that the electronic and lattice heat capacities $C_e,C_l$ in principle depend on their respective temperatures $T_e,T_l$ so that the extended TTM model is nonlinear. However, in the limit $\Delta T_e(t) << T_{eq}$, it is possible to approximate $C_e,C_l$ as independent of their respective temperatures \cite{Carpene_PRB_2006}. Setting $\partial_t N = 0$ and removing the equation for $N(t)$ is equivalent to neglecting the thermalization time $\tau_{th}$ over which non-thermalized electrons release energy to the thermalized ones. This characteristic time is of order $\tau_{th}\approx 300 fs$ and can be neglected for long pulses. However, for pulses of duration $\tau_0 \approx 100 fs$ such an approximation is not feasible and all the equations must be retained in order to describe correctly the delayed nonlinearity. The TTM model can be solved straightforwardly in the Fourier domain, leading to the solution
\begin{eqnarray}
\Delta T_e (\Delta\omega) & = & T_e (\Delta\omega) - T_l (\Delta\omega) =   \\
                        & = & \tau_r\tau_{th}\left(\frac{\gamma_e}{C_e}-\frac{\gamma_l}{C_l}\right) h_T(\Delta\omega) P_A(\Delta\omega) , \nonumber
\end{eqnarray}
where $\Delta\omega$ is the shift from the carrier angular frequency $\omega_0$ of the ultrashort optical pulse and
\begin{equation}
h_T(\Delta\omega) =  \frac{ 1 }{ [ 1 - i \tau_{th} \Delta\omega ] [ 1 - i \tau_r \Delta\omega ] }.
\end{equation}

\begin{figure}
\centering
\begin{center}
\includegraphics[width=0.45\textwidth]{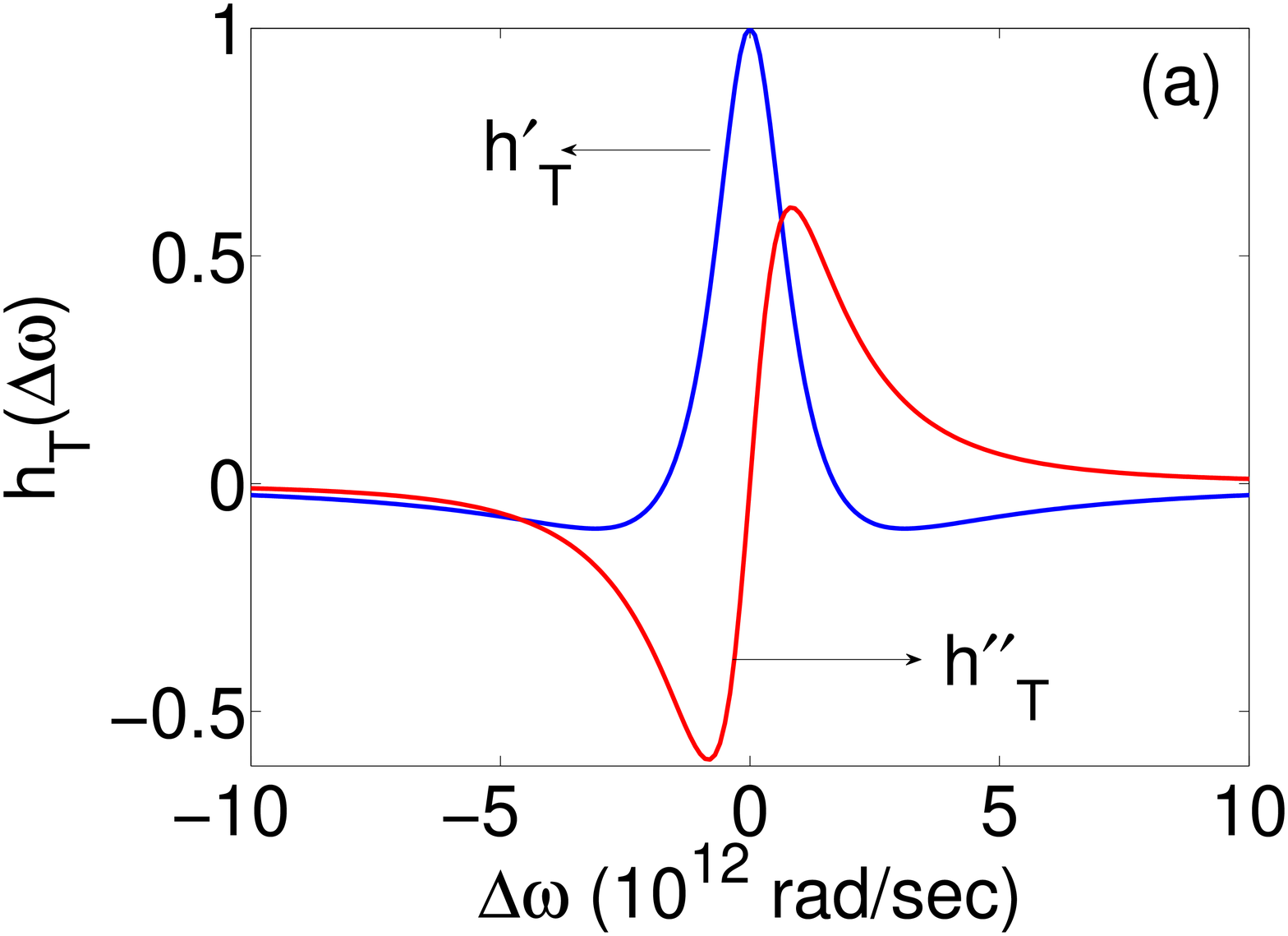}
\includegraphics[width=0.45\textwidth]{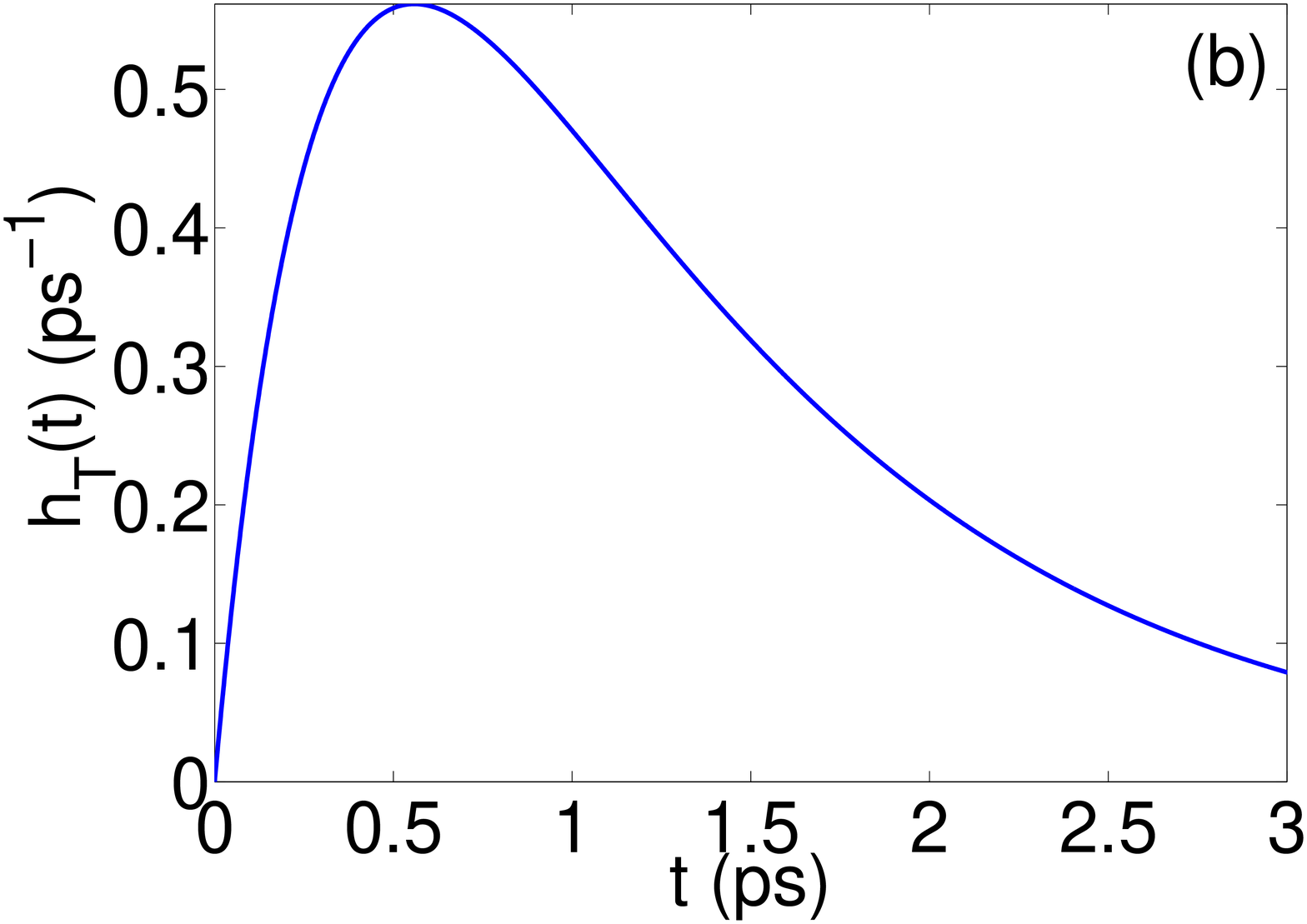}
\caption{Plot of the two-temperature model (TTM) response. (a) Fourier transform $h_T(\Delta\omega)$ as a function of angular frequency shift $\Delta\omega$. The blue and red curves represent the real ($h'_T(\Delta\omega)$) and imaginary ($h''_T(\Delta\omega)$) parts. (b) TTM temporal response $h_T(t)$ as a function of time $t$.}
\label{TTM_Resp_Fun}
\end{center}
\end{figure}

In the expressions above we have used the parameter $\tau_r = C_e C_l/[{\cal C}(C_e+C_l)]$ representing the relaxation time of the thermalized electrons with the lattice. In the temporal domain, the temperature variation $\Delta T_e(t) = T_e(t) - T_l(t)$ is given by
\begin{eqnarray}
&& \Delta T_e (t) = \frac{1}{2\pi}\int_{-\infty}^{+\infty}d\Delta\omega \Delta T (\Delta\omega)e^{-i\Delta\omega t} = \\
&& = \tau_r\tau_{th}\left(\frac{\gamma_e}{C_e}-\frac{\gamma_l}{C_l}\right)\int_{-\infty}^{+\infty} h_T(t') P_A(t-t') dt' , \nonumber
\end{eqnarray}
where the temporal response function $h_T(t)$ is
\begin{equation}
h_T(t) =  \frac{\theta(t)}{\tau_{th}-\tau_r}  \left( e^{-t/\tau_{th}} - e^{- t/\tau_r} \right). \label{THRNMMODINTRBNDNLRSPFNCTN}
\end{equation}
Note that causality is imposed by means of the Heaviside step function $\theta(t)$. Note also that in the limit $P_A \rightarrow 0$ the conduction electrons and the lattice are in equilibrium and have the same temperature $T_e = T_l = T_{eq}$. Since the heat capacity of the lattice $C_l$ is much greater than the electronic heat capacity $C_e$, the temperature difference $\Delta T_e(t) = T_e(t) - T_l(t)$ can be approximated by $\Delta T_e(t) \approx T_e(t) - T_{eq}$. The response functions in the frequency $h_T(\Delta\omega)$ and temporal $h_T(t)$ domains are plotted as functions of $\Delta\omega$ and $t$ in Figs. \ref{TTM_Resp_Fun}a,b. In Fig. \ref{TTM_Resp_Fun}a, the blue and red curves correspond to the real $h'_T(\Delta\omega)$ and imaginary $h''_T(\Delta\omega)$ parts. Note that, following the sign convention chosen for the exponential in the Fourier expansion ($e^{-i\Delta\omega t}$), a positive value of the imaginary part $h''_T(\Delta\omega)$ corresponds to loss, while a negative value corresponds to gain. The temporal thermal response $h_T(t)$, depicted in  Fig. \ref{TTM_Resp_Fun}b,  is mainly characterized by a peak delayed in time by $\Delta t \approx 600 fs$. As a consequence of the delay, blue-shifted frequency components are suppressed, while red-shifted components are amplified, analogously to what happens in solid-core optical fibers as a result of the Raman effect \cite{Agrawal2001}.

\subsection{Thermo-modulational interband nonlinear susceptibility}

In the previous sections we have described the temperature dependence of the {\it linear} dielectric function of gold and the temporal dynamics of the conduction electrons heated by an ultrashort optical pulse. In this section we sum up the results obtained so far calculating the {\it nonlinear} susceptibility due to Fermi smearing of the conduction electrons. The instantaneous power per unit volume absorbed by a metal is $W(t)=\vec{E}(t)\cdot\partial_t\vec{D}(t)$, where
\begin{equation}
\vec{D}(t) = \epsilon_0 \int_{-\infty}^{+\infty} \epsilon_m(t-t')\vec{E}(t'),
\end{equation}
$\epsilon_0$ is the vacuum permittivity and $\epsilon_m(t-t')$ is the temporal dielectric response function. In the continuous wave (CW) monochromatic case, the electric and displacement fields are $\vec{E}(t) = \vec{E}_0 e^{-i\omega_0 t}$ and $\vec{D}(t) = \epsilon_0 \epsilon_m(\omega_0) \vec{E}_0 e^{-i\omega_0 t}$, where $\omega_0$ is the angular frequency. The mean absorbed power can be calculated by averaging the instantaneous power over the fast oscillations $e^{-i\omega_0 t}$ \cite{Landau_Book}: $P_A = (1/2)\epsilon_0 \omega_0 \epsilon''_m(\omega_0) |\vec{E}|^2$. Hence, inserting the power dependent temperature variation $\Delta T_e$ into Eq. (\ref{ThermoModDelteps}), the nonlinear polarization is given by
\begin{equation}
\vec{P}_{NL}^{CW} (t) = \epsilon_0 \chi^{(3)}_T(\omega_0) |\vec{E}|^2 \vec{E}(t).
\end{equation}
$\chi^{(3)}_T(\omega_0)$ is the {\it thermo-modulational interband nonlinear susceptibility}:
\begin{equation}
\chi^{(3)}_T(\omega_0) = \frac{1}{2} \epsilon_0 \omega_0 \epsilon''_m(\omega_0) \gamma_T (\omega_0) , \label{ThrmModNLSuscEqChi3}
\end{equation}
where
\begin{equation}
\gamma_T (\omega) = \tau_r\tau_{th} \left(\frac{\gamma_e}{C_e}-\frac{\gamma_l}{C_l}\right) \partial_{T_e} \epsilon_{inter}(\omega)  \label{ThrmModNLSuscEqGM}.
\end{equation}

\begin{figure}
\centering
\begin{center}
\includegraphics[width=0.45\textwidth]{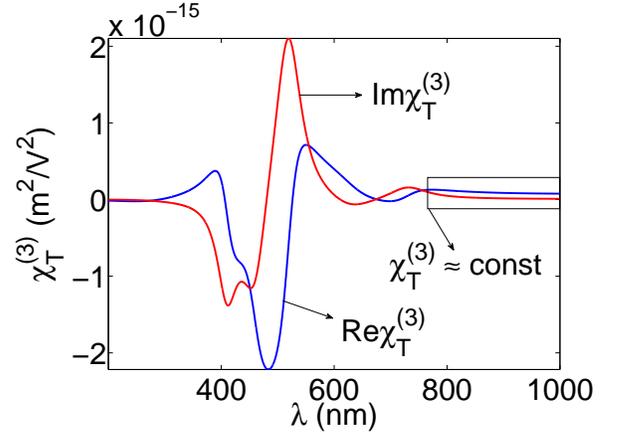}
\caption{Thermo-modulational interband nonlinear susceptibility $\chi_T^{(3)}$ as a function of the optical wavelength $\lambda$. Blue and red curves correspond to the real and imaginary parts of $\chi^{(3)}_T(\lambda)$. The region within which $\chi_T^{(3)}$ can be approximately considered to be a constant is also indicated.}
\label{Chi3_Thrm_Mod_Fig}
\end{center}
\end{figure}

The real and imaginary parts of $\chi^{(3)}_T$ are plotted as functions of optical wavelength $\lambda$ in Fig. \ref{Chi3_Thrm_Mod_Fig}. Note that, as a consequence of the resonant interband transitions, the nonlinear susceptibility is strongly dispersive at optical frequencies and can be much greater ($\approx 7$ orders of magnitude) than the Kerr susceptibility of bulk silica ($\chi_3^{Si} \approx 10^{-22} m^2/V^2$). The strong frequency dispersion of gold dramatically changes its optical properties, as well as the signs of ${\mathrm Re}\chi^{(3)}_T,{\mathrm Im}\chi^{(3)}_T$. Note that, for wavelengths $\lambda \gtrsim 750 nm$, the thermo-modulational nonlinear susceptibility can be approximated as $\chi^{(3)}_T\approx const$.

For ultrashort optical pulses, the calculation of the nonlinear dielectric polarization is more involved. In the slowly varying envelope approximation (SVEA) the electric field can be expressed as $\vec{E}(t) = \psi(t)e^{ - i \omega_0 t}\hat{n}$, where $\omega_0$ is the carrier angular frequency, $\hat{n}$ is the polarization unit vector and $\psi(t)$ is the envelope amplitude, which is slowly varying compared to the fast oscillations $e^{-i\omega_0 t}$. The expression for the mean absorbed power $P_A(t)$ in this non-monochromatic case includes also the contributions of first-order dispersion \cite{Landau_Book} and is explicitly given by
\begin{eqnarray}
&& P_A(t) = \frac{\epsilon_0}{4} \left\{ 2\omega_0\epsilon''_m(\omega_0)|\psi|^2 + \left.\frac{d(\omega\epsilon'_m)}{d\omega}\right|_{\omega_0}\partial_t|\psi|^2 + \right. \nonumber \\
&& \left. + i\left.\frac{d(\omega\epsilon''_m)}{d\omega}\right|_{\omega_0} (\psi^*\partial_t\psi-\psi\partial_t\psi^*) \right\}. \label{AbsPowTotExpr}
\end{eqnarray}
In conclusion, the nonlinear polarization created by an ultrafast optical pulse can be expressed in terms of a double convolution integral
\begin{eqnarray}
\vec{P}_{NL} (t) & = & \epsilon_0 \int_{0}^{+\infty} dt' \int_{0}^{+\infty} dt'' \times \label{NL_POL_AU_DEF_EQ} \\
                 &   & \times \gamma_T (t') h_T(t'') P_A(t-t'-t'') \vec{E}(t-t'). \nonumber
\end{eqnarray}
In this expression, $\gamma_T (t)$ is the interband response function (measured in the units $m^3W^{-1}s^{-1}$) and is given by the inverse Fourier transform of $\gamma_T (\omega)$ (measured in the units $m^3W^{-1}$), which is in turn given by Eq. (\ref{ThrmModNLSuscEqGM}).

\section{Thermo-modulational nonlinear dynamics in plasmonic devices}

In the previous sections we have calculated the thermo-modulational interband nonlinear susceptibility of gold starting from the basic properties of its band structure. In this section we study the optical propagation of surface plasmon polaritons (SPPs) guided along gold nanowires surrounded by silica glass, including the novel nonlinear effects originating from the heating of gold. A common theoretical approach to modelling optical propagation in optical fibres and plasmonic waveguides uses the nonlinear Schr\"odinger equation for the slowly varying amplitude of a guided pulse perturbatively derived based on the assumptions of low loss and nonlinearity \cite{Agrawal2001,Afshar_OE_2009,Afshar_OL_2009,Tran_OE_2009,Biancalana_PRL_2010,Daniel_JOSAB_2010,Skryabin_JOSAB_2011,Marini_PRA_2011}.

\begin{figure}
\centering
\begin{center}
\includegraphics[width=0.3\textwidth]{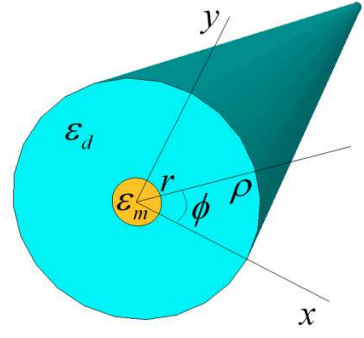}
\caption{Gold nanowire of radius $r$ surrounded by silica glass.}
\label{Metallic_Nanowire_Geometry_Fig}
\end{center}
\end{figure}

We start the analysis from the time-dependent Maxwell equations for the optical electric ($\vec{E}$) and magnetic ($\vec{H}$) fields:
\begin{eqnarray}
\nabla \times \vec{E}(\vec{r},t) & = & - \mu_0\partial_t\vec{H}(\vec{r},t), \label{CurlE}\\
\nabla \times \vec{H}(\vec{r},t) & = & \partial_t \vec{D}_L(\vec{r},t) + \partial_t \vec{P}_{NL}(\vec{r},t), \label{CurlH}
\end{eqnarray}
where $\vec{r}$ is the position vector, $t$ is the temporal variable, $\mu_0$ is the magnetic permeability of vacuum, $\vec{P}_{NL}(\vec{r},t)$ is the nonlinear dielectric polarization and $\vec{D}_L(\vec{r},t)$ is the linear part of the displacement vector, which is given by the constitutive relation
\begin{equation}
\vec{D}_L (\vec{r},t) = \epsilon_0 \int_{-\infty}^{+\infty} dt' \epsilon_L(\vec{r},t') \vec{E}(\vec{r},t-t').
\end{equation}
$\epsilon_L(\vec{r},t)$ is the position-dependent temporal response function, which can be expressed in terms of a Fourier expansion of the linear dielectric profile $\epsilon_L(\vec{r},\omega)$:
\begin{equation}
\epsilon_L(\vec{r},t) = \frac{1}{2\pi}\int_{-\infty}^{+\infty} d\omega \epsilon_L(\vec{r},\omega) e^{-i\omega t}.
\end{equation}
As a consequence of the cylindrical symmetry of the gold nanowire around the $z$-axis, the dielectric profile depends solely on the modulus of the position vector $\rho=|\vec{r}|$:
\begin{equation}
\epsilon_L(\rho,\omega) = \epsilon_m(\omega)\theta(r-\rho) + \epsilon_d(\omega)\theta(\rho-r),
\end{equation}
where $\theta(x)$ is the Heaviside step function, $\omega$ is the angular frequency, $r$ is the radius of the nanowire and $\epsilon_m(\omega),\epsilon_d(\omega)$ are the linear dielectric constants of gold and silica. In the calculations below we use the Sellmeier expansion for the dielectric constant of silica $\epsilon_d(\omega)$ and a Lorentzian fit to the experimental data for the linear dielectric constant of gold $\epsilon_m(\omega)$, as reported in Ref.  \cite{Ung2007}. A sketch of the plasmonic structure discussed in this section is depicted in Fig. \ref{Metallic_Nanowire_Geometry_Fig}.

\subsection{Linear modes}

If one neglects the nonlinear polarization $\vec{P}_{NL}$, Eqs. (\ref{CurlE},\ref{CurlH}) can be directly solved using the Ansatz:
\begin{eqnarray}
\vec{E}(\vec{r},t) & = & I^{1/2}\psi \vec{e}(\rho) e^{im\phi + i\beta z - i\omega t}, \label{ansatz_E} \\
\vec{H}(\vec{r},t) & = & \epsilon_0 c I^{1/2}\psi\vec{h}(\rho) e^{im\phi + i\beta z - i\omega t}, \label{ansatz_H}
\end{eqnarray}
where $c$ is the speed of light in vacuum, $\psi$ is the mode amplitude (measured in $W^{1/2}$), $\vec{e},\vec{h}$ are the linear guided mode profiles (dimensionless), $\beta$ is the mode propagation constant, $\phi$ is the angle between the vectors $\vec{r},\hat{x}$ and $m$ is the azimuthal mode order. The factor $I^{1/2}$ is a constant, chosen in such a way that $|\psi|^2$ represents the total optical power carried by a linear mode with angular frequency $\omega$ and propagation constant $\beta$. The linear dispersion relation $\beta(\omega)$ can be directly calculated by substituting Eqs. (\ref{ansatz_E},\ref{ansatz_H}) into Eqs. (\ref{CurlE},\ref{CurlH}) and by applying the boundary conditions for the continuity of the tangential components of the electric field and the normal component of the displacement vector \cite{Marcuse_Book,Novotny_PRE_1994,Takahara_OL_1997,Schmidt_OE_2008,Marini_PRA_2011}. The modal profiles $\vec{e},\vec{h}$ are combinations of modified Bessel and Hankel functions of different orders \cite{Marcuse_Book,Novotny_PRE_1994,Takahara_OL_1997,Schmidt_OE_2008,Marini_PRA_2011}, the solutions $\vec{E}(\vec{r},t),\vec{H}(\vec{r},t)$ corresponding to SPPs \cite{Maier_Book}.

\begin{figure}
\centering
\begin{center}
\includegraphics[width=0.45\textwidth]{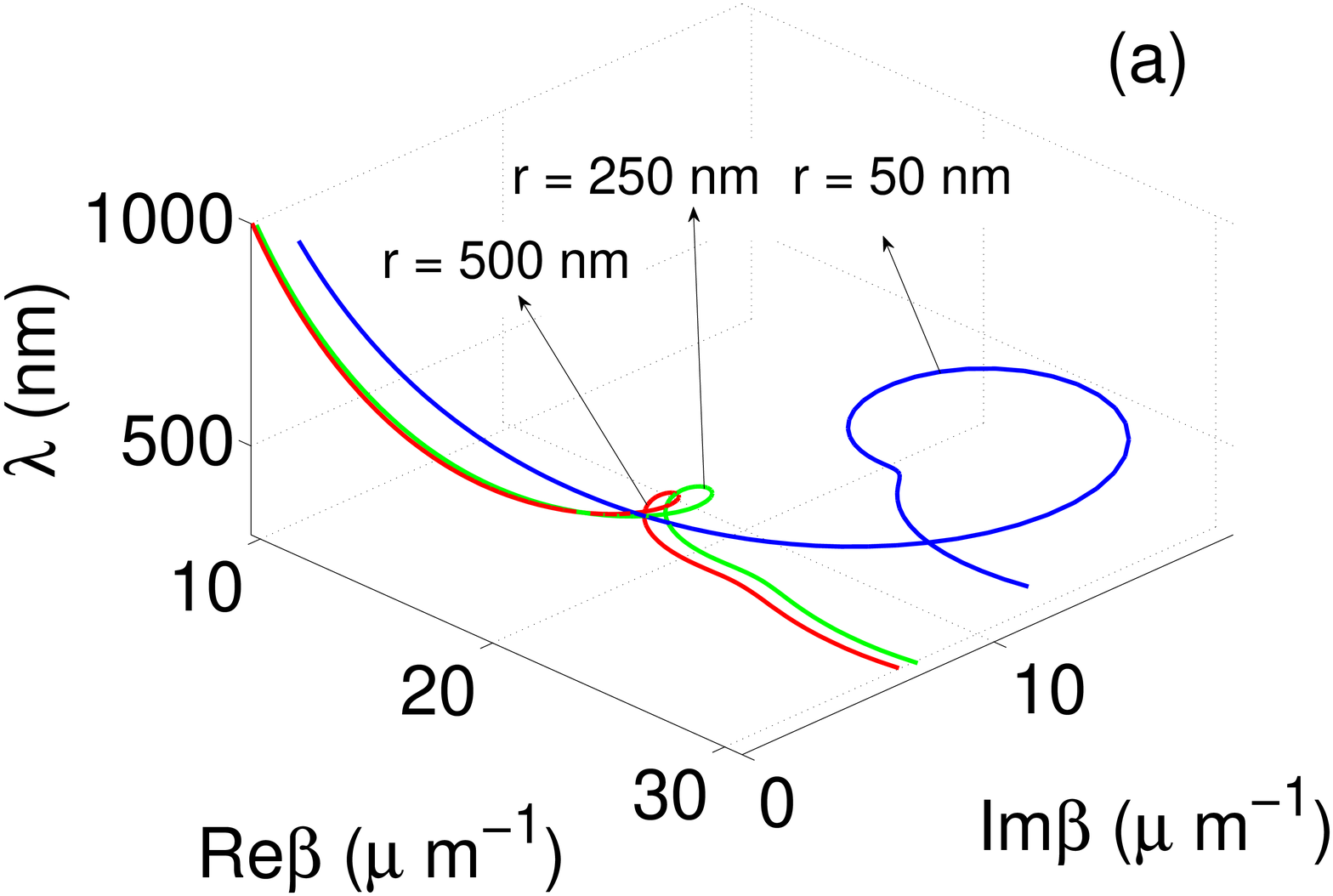}
\includegraphics[width=0.45\textwidth]{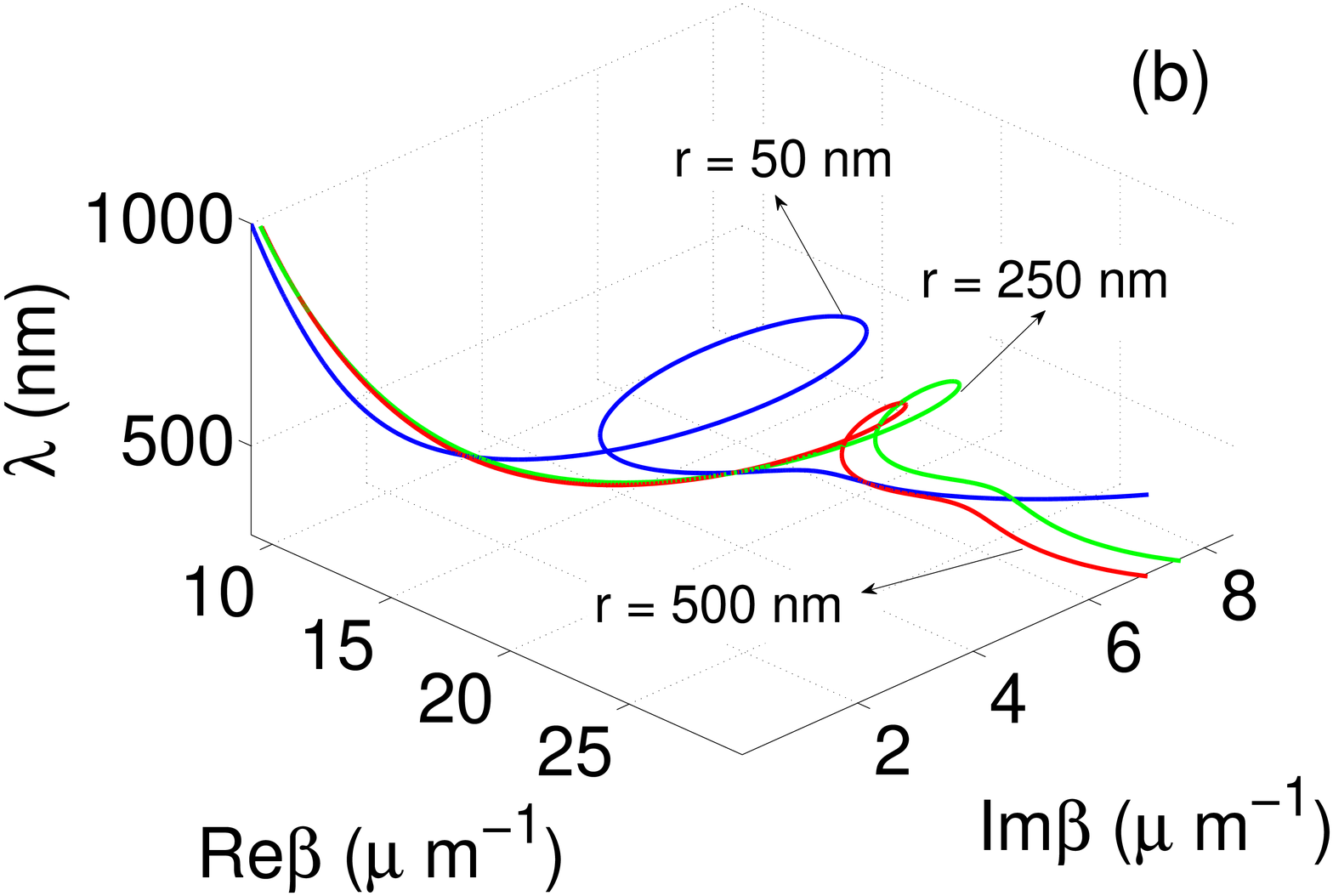}
\caption{Linear dispersion relations $\beta(\lambda)$ for the (a) $m=0$ and (b) $m=1$ plasmon polariton modes. Blue, green and red curves correspond to the wire radii $r=50,250,500 ~ nm$, respectively.}
\label{Linear_Beta_3D_Fig}
\end{center}
\end{figure}

Figs. \ref{Linear_Beta_3D_Fig}a,b show the complex dispersion relations for the (a) $m=0$ and (b) $m=1$ guided SPP modes; real and imaginary parts of the propagation constant $\beta$ are plotted as functions of the optical wavelength $\lambda$. For both the $m=0,1$ modes, if $\lambda\gtrsim 500 nm$, the real (${\mathrm Re}\beta$) and imaginary (${\mathrm Im}\beta$) parts of the propagation constant increase as the optical wavelength $\lambda$ decreases, reaching maxima at $\lambda = \lambda_{sp} \simeq 500 nm$, at the surface plasmon resonance. For $\lambda \lesssim 500 nm$, the behaviour of the dispersion relation is more involved. In the ideal case where the metal is lossless, the imaginary part of the propagation constant vanishes, ${\mathrm Im}\beta=0$, while the real part ${\mathrm Re}\beta$ diverges as $\lambda\rightarrow\lambda_{sp}$.

\begin{figure}
\centering
\begin{center}
\includegraphics[width=0.44\textwidth]{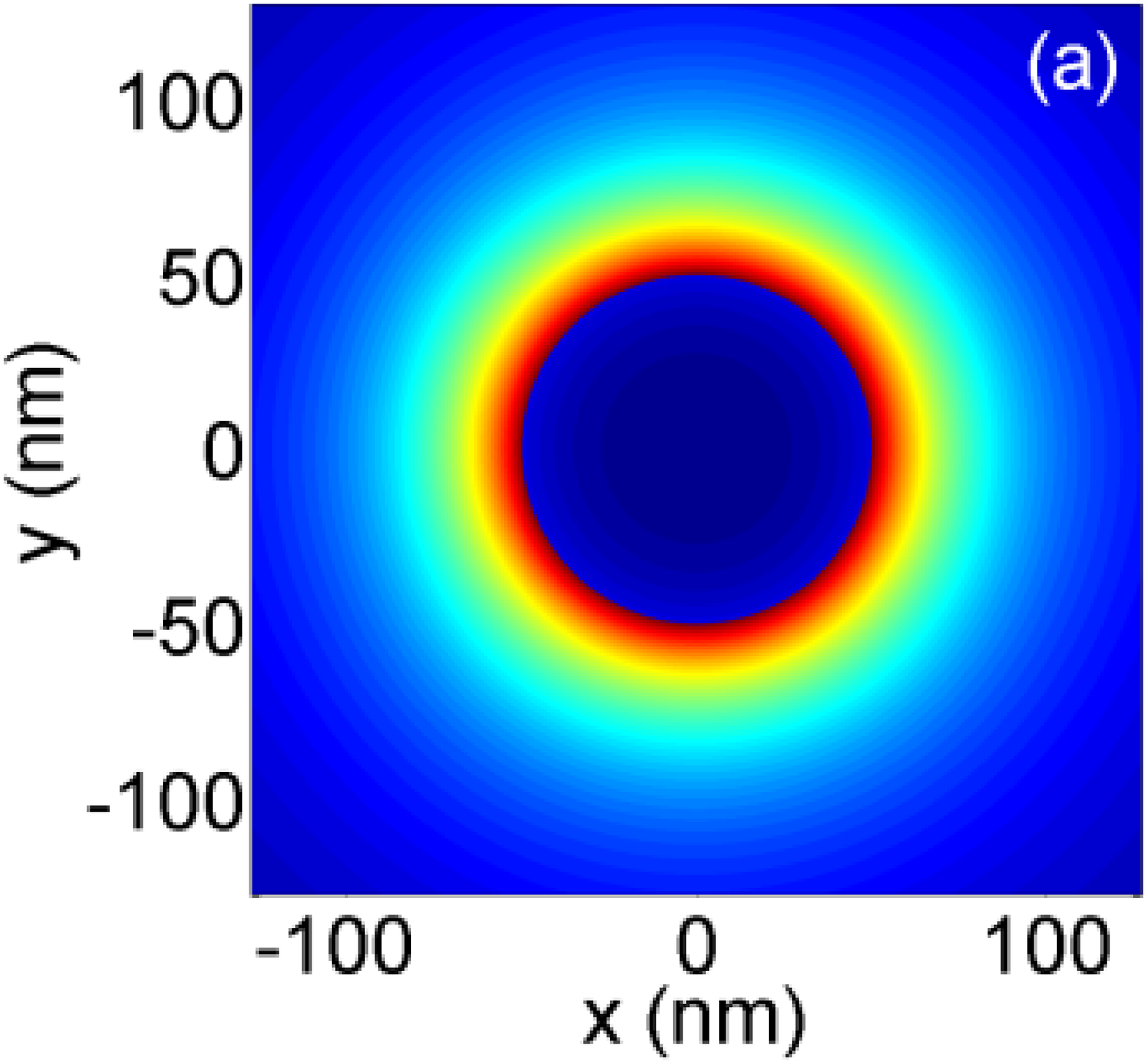}
\includegraphics[width=0.44\textwidth]{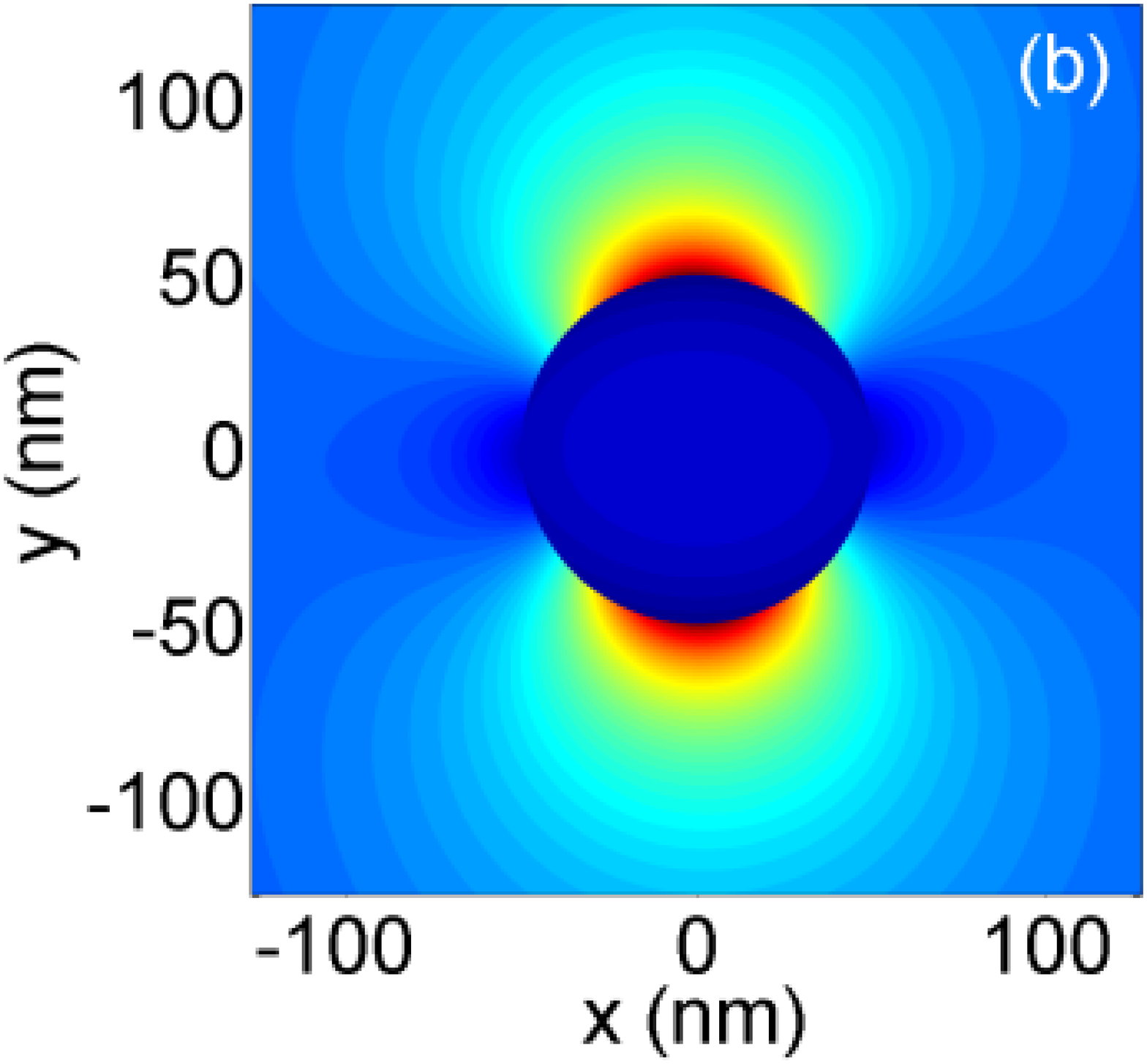}
\caption{Contour-plots of the time-averaged Poynting vector for the (a) $m=0$ mode and (b) a superposition of the $m = \pm 1$ modes on a gold nanowire with radius $r=50 nm$ surrounded by silica. The optical wavelength is $\lambda = 800 nm$.}
\label{Sz_Cntrplt_m_0_Fig}
\end{center}
\end{figure}

Optical confinement depends mainly on the real part of the propagation constant ${\mathrm Re}\beta$: high values of ${\mathrm Re}\beta$ correspond to tightly confined modes \cite{Marcuse_Book,Schmidt_OE_2008}. On the other hand, the attenuation coefficient of the linear modes is directly related to the imaginary part of the propagation constant: $\alpha = 2 {\mathrm Im}\beta$. Hence, if one wants to achieve tight SPP confinement it is impossible to avoid high losses so that the use of materials with large gain is required in practical applications \cite{Bergman2003,Protsenko_PRA_2005,Stockman_NatPhot_2008,Zheludev2008,Oulton_NatLett_2009,Marini2009,Gather_Nature_2010}. Note that the fundamental mode $m=0$ is TM polarized and that both the real and imaginary parts of the propagation constant ${\mathrm Re}\beta,{\mathrm Im}\beta$ increase as the radius of the gold nanowire decreases (see Fig. \ref{Linear_Beta_3D_Fig}a). A contour-plot of the time-averaged Poynting vector $S_z = (1/2) \hat{z} \cdot Re( \vec{E} \times \vec{H}^* )$ of the TM plasmonic mode ($m=0$) of a gold nanowire with radius $r = 50 nm$ at optical wavelength $\lambda = 800nm$ is depicted in Fig. \ref{Sz_Cntrplt_m_0_Fig}a. Note that the electromagnetic field is tightly bound to the metal surface and that the power distribution is cylindrically symmetric.

In contrast to the TM fundamental mode, the $m=1$ mode is hybrid polarized and less well confined. The dispersion relation does not depend on the sign of the azimuthal mode order $m$ \cite{Marcuse_Book} so that the $m = \pm 1$ modes (characterized by opposite chirality) are degenerate. A contour-plot of the time-averaged Poynting vector $S_z$ of the superposition of $m = \pm 1$ guided SPP modes is shown in Fig. \ref{Sz_Cntrplt_m_0_Fig}b (for the same parameters of Fig. \ref{Sz_Cntrplt_m_0_Fig}a). 

\begin{figure}
\centering
\begin{center}
\includegraphics[width=0.45\textwidth]{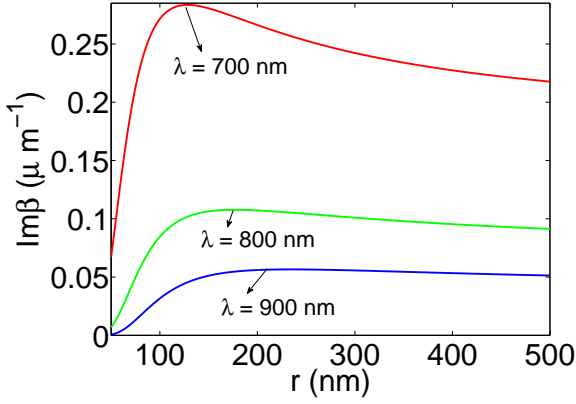}
\caption{Imaginary part of the propagation constant $Im\beta$ as a function of the wire radius $r$ for the hybrid $m=1$ plasmonic mode. Red, green and blue curves correspond to the optical wavelengths $\lambda = 700,800,900 ~ nm$, respectively.}
\label{Beta_Im_R_TOT_m_1_Fig}
\end{center}
\end{figure}

Note that the power distribution of such mode is not azimuthally symmetric and depends on the angle $\phi$. If one calculates the time-dependent Poynting vector of the $m = \pm 1$ modes, finds that these SPPs spiral around the surface of the gold nanowire with opposite chirality \cite{Schmidt_OE_2008}. However, the azimuthal spiralling is averaged out in time and as a result there is no net angular flow of time-averaged power for the single $m=\pm 1$ modes and for their superposition. In Fig. \ref{Beta_Im_R_TOT_m_1_Fig},  ${\mathrm Im}\beta$ for the $m=1$ mode is plotted as a function of $r$. Red, green and blue curves correspond to $\lambda = 700,800,900 ~ nm$. Both ${\mathrm Re}\beta$ and ${\mathrm Im}\beta$ depend on $r$ in a manner more complicated than for the fundamental TM mode (see Figs. \ref{Linear_Beta_3D_Fig}a,b,\ref{Beta_Im_R_TOT_m_1_Fig}). In particular, for fixed optical wavelength $\lambda$, ${\mathrm Im}\beta$ is maximum at a characteristic wire radius $r_0$, decreasing significantly when $r<r_0$. Hence, if the wire radius is much smaller than the optical wavelength $r\ll\lambda$, {\it long range surface plasmon polaritons} can be excited \cite{Schmidt_OE_2008,Maier_Book}. The field penetration within the gold nanowire is limited for these modes, and hence the attenuation is significantly reduced. In the following nonlinear analysis we focus on the $m=0,1$ modes. Higher order modes are less well confined and cut off for wavelengths greater than a particular value $\lambda_{co}$.

\subsection{Generalized nonlinear Schr\"odinger equation}

If the effect of loss and nonlinearity on the fast linear oscillations is weak, the nonlinear propagation of an optical pulse in a plasmonic waveguide can be described by the generalized nonlinear Schr\"odinger equation (GNLSE) for the field amplitude in the slowly varying envelope approximation (SVEA) \cite{Agrawal2001,Afshar_OE_2009,Afshar_OL_2009,Tran_OE_2009,Biancalana_PRL_2010,Daniel_JOSAB_2010,Skryabin_JOSAB_2011,Marini_PRA_2011}. In this approach the Ansatz for the electromagnetic field is
\begin{eqnarray}
\vec{E}(\vec{r},t) & = & I^{1/2}\psi(z,t) \vec{e}(\rho) e^{im\phi + i\beta_0 z - i\omega_0 t}, \label{ansatz_E_NL} \\
\vec{H}(\vec{r},t) & = & \epsilon_0 c I^{1/2}\psi(z,t)\vec{h}(\rho) e^{im\phi + i\beta_0 z - i\omega_0 t}, \label{ansatz_H_NL}
\end{eqnarray}
where $\psi(z,t)$ is the slowly varying envelope amplitude and $\omega_0$ is the carrier angular frequency. With $\beta_0$ we denote the linear propagation constant at the carrier frequency calculated by neglecting the nonlinear polarization and the metal loss; $\vec{e}(\rho),\vec{h}(\rho)$ are the corresponding unperturbed linear mode profiles (dimensionless) and $I$ is a constant chosen so that $|\psi|^2$ represents the optical power, as in the previous section. For a gold nanowire surrounded by silica glass, the nonlinear polarization is
\begin{equation}
\vec{P}_{NL}(\vec{E}) = \vec{P}_{NL}^{Au}(\vec{E})\theta(r-\rho) + \vec{P}_{NL}^{Si}(\vec{E})\theta(\rho-r),
\end{equation}
where
\begin{equation}
\vec{P}_{NL}^{Si}(\vec{E}) = \frac{\epsilon_0}{2}\chi_{Si}^{(3)}\left[|\vec{E}|^2\vec{E}+\frac{1}{2}\vec{E}^2\vec{E}^*\right], \label{Silica_NL_POL_EQ}
\end{equation}
$\chi_{Si}^{(3)} = 2.25 \times 10^{-22} m^2/V^2$ is the Kerr coefficient of silica glass at $\lambda_0 = 800 nm $ and $\vec{P}_{NL}^{Au}(\vec{E})$ is given by Eq. (\ref{NL_POL_AU_DEF_EQ}). Note that in Eq. (\ref{Silica_NL_POL_EQ}) we have neglected the Raman effect \cite{Blow_IEEE_1989}, which should in principle be retained. However, as we will show, the effective nonlinear coefficient of silica is much smaller than the effective nonlinear coefficient of gold so that neither Kerr nor Raman effects play any significant role. Hence, for the sake of simplicity, we do not consider the Raman term, comparing our results only with the Kerr term.

Since we are mainly interested in the thermo-modulational interband nonlinearity, we neglect the dispersive terms in the absorbed power $P_A(t)$, given by Eq. (\ref{AbsPowTotExpr}). These terms are expected to play only a minor role, since they are small corrections to the carrier term. In what follows, we will focus on the spectral region ($\lambda \approx 800 nm$) where the interband response function is approximately constant ($\gamma_T(\omega)\approx\gamma_T(\omega_0)$, see Fig. \ref{Chi3_Thrm_Mod_Fig}) so that the nonlinear polarization of gold can be approximated by
\begin{equation}
\vec{P}_{NL}^{Au}(\vec{E}) \approx \epsilon_0 \chi^{(3)}_{Au} (\omega_0) \int_{0}^{+\infty} dt' h_T(t')|\vec{E}(t-t')|^2 \vec{E}(t),
\end{equation}
where $h_T(t'),\chi^{(3)}_{Au} (\omega_0)$ are given by Eqs. (\ref{THRNMMODINTRBNDNLRSPFNCTN},\ref{ThrmModNLSuscEqChi3}). By inserting Eqs. (\ref{ansatz_E_NL},\ref{ansatz_H_NL}) into Maxwell equations and developing a first order perturbative theory \cite{Marini_PRA_2011} one obtains the GNLSE for the slowly varying amplitude $\psi(z,t)$:
\begin{eqnarray}
&& i\partial_z \psi(z,t) + \hat{D}(i\partial_t) \psi(z,t) + \Upsilon_{Si} |\psi(z,t)|^2\psi(z,t) + \nonumber \\
&& + \Upsilon_{Au} \int_{0}^{+\infty} dt' h_T(t') |\psi(z,t-t')|^2 \psi(z,t) = 0, \label{PropEq}
\end{eqnarray}
where
\begin{eqnarray}
&& \Upsilon_{Si} = \frac{\omega_0\chi_{Si}^{(3)}}{4\epsilon_0 c^2} \frac{\int_0^{2\pi}d\phi \int_r^{+\infty}d\rho \rho  \left[2|\vec{e}|^4+|\vec{e}^2|^2\right]}{\left(\int_0^{2\pi}d\phi\int_0^{+\infty}d\rho\rho Re\left[\vec{e}\times\vec{h}^*\right]\cdot\hat{z}\right)^2 }, \nonumber \\
&& \Upsilon_{Au} = \frac{\omega_0\chi^{(3)}_{Au}}{\epsilon_0 c^2} \frac{\int_0^{2\pi}d\phi \int_0^rd\rho \rho |\vec{e}|^4 }{\left(\int_0^{2\pi}d\phi\int_0^{+\infty}d\rho\rho Re\left[\vec{e}\times\vec{h}^*\right]\cdot\hat{z}\right)^2 }. \nonumber
\end{eqnarray}
The linear dispersion operator $\hat{D}(i\partial_t)$ is complex, accounting as it does for the linear losses of gold. Its action on the envelope amplitude can be calculated in the Fourier domain:
\begin{equation}
\hat{D}(i\partial_t) \psi(z,t) = \frac{1}{2\pi}\int_{-\infty}^{+\infty}d\omega D(\omega)\psi(z,\omega)e^{-i\omega t},
\end{equation}
where
\begin{equation}
D(\omega) = \beta(\omega)-\beta_0 - \left.\frac{d\beta'}{d\omega}\right|_{\omega_0} (\omega-\omega_0).
\end{equation}
Note that $\beta_0$ is the real-valued carrier propagation constant of the linear unperturbed mode, $\beta(\omega)$ is the complex modal wavevector calculated in the previous section and the prime superscript in the equation above indicates the real part ($\beta'={\mathrm Re}\beta$).

\begin{figure}
\centering
\begin{center}
\includegraphics[width=0.45\textwidth]{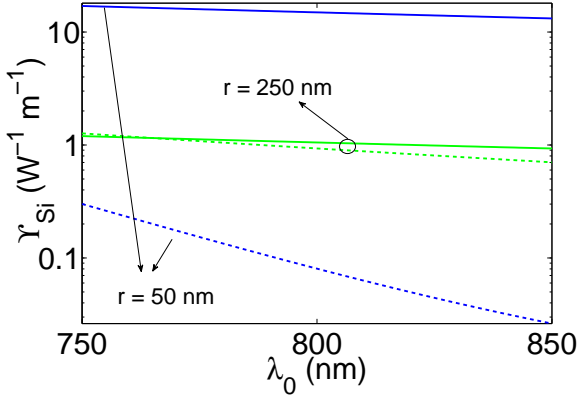}
\caption{Kerr nonlinear coefficient ($\Upsilon_{Si}$) of a gold nanowire surrounded by silica glass for the $m=0$ (full lines) and $m=1$ (dashed lines) modes. Blue and green lines correspond to the wire radii $r=50,250 ~ nm$, respectively. Note that the plot is made in semi-logarithmic scale.}
\label{Kerr_NL_COEFF_Fig}
\end{center}
\end{figure}

\begin{figure}
\centering
\begin{center}
\includegraphics[width=0.45\textwidth]{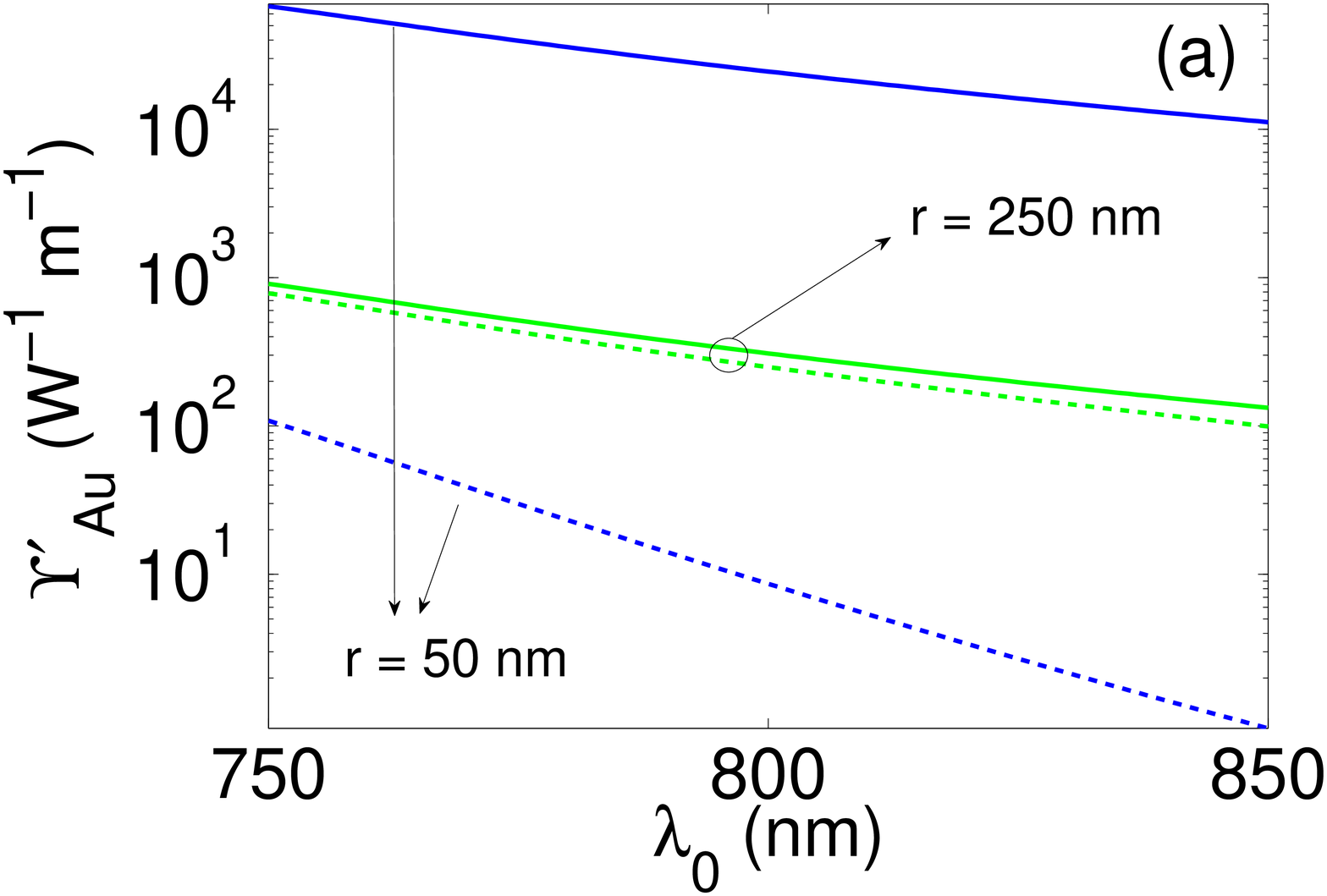}
\includegraphics[width=0.45\textwidth]{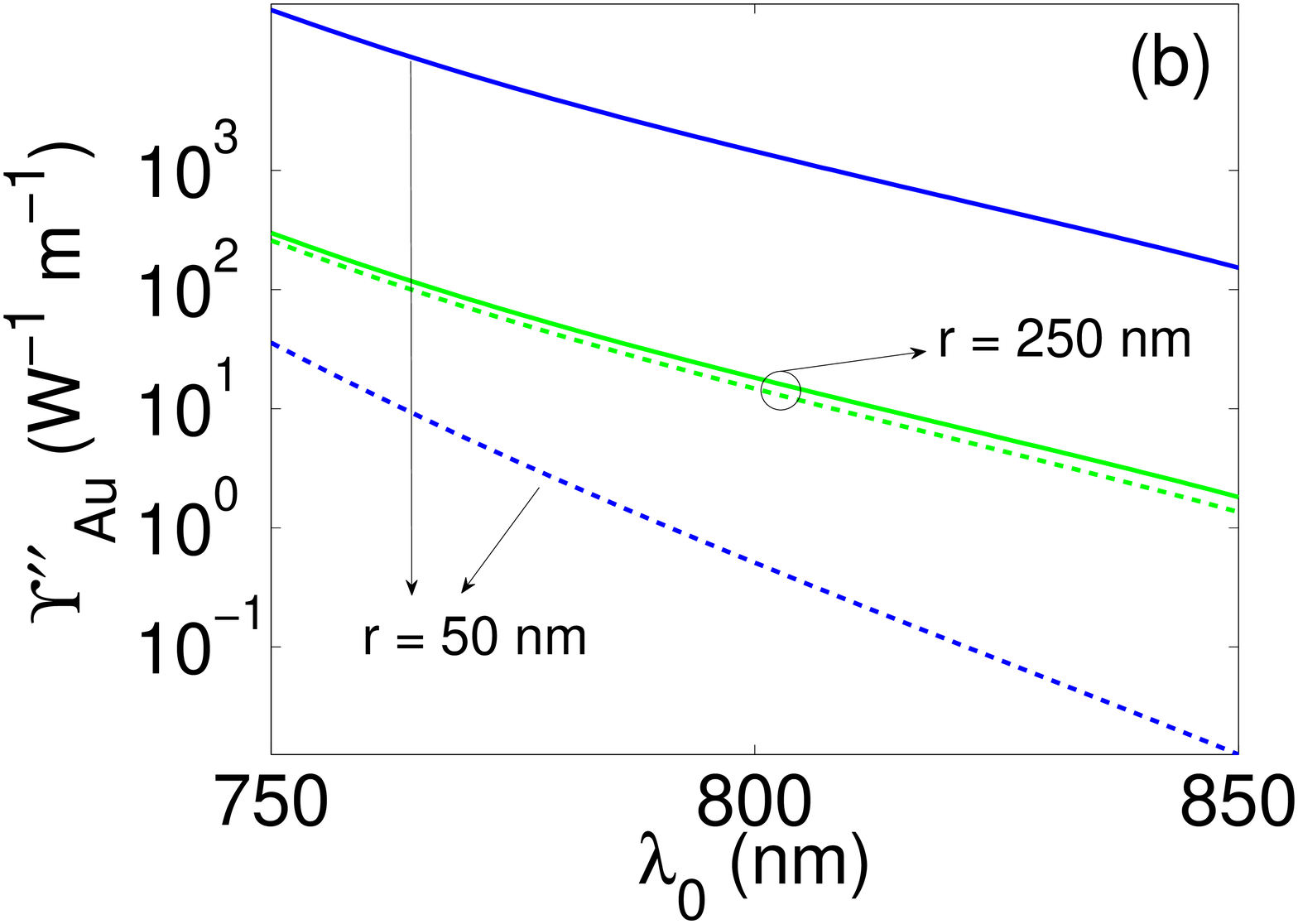}
\caption{(a) Real and (b) imaginary parts of the thermo-modulational interband nonlinear coefficient ($\Upsilon_{Au}$) of a gold nanowire surrounded by silica glass for $m=0$ (full curves) and $m=1$ (dashed curves) modes. Blue and green curves correspond to the wire radii $r=50,250 ~ nm$, respectively. Note that the plots are made in semi-logarithmic scale.}
\label{Gold_NL_COEFF_Fig_m_both}
\end{center}
\end{figure}

The nonlinear parameters $\Upsilon_{Si},\Upsilon_{Au}$ are measured in $m^{-1}W^{-1}$ and account also for the surface nonlinearity \cite{Skryabin_JOSAB_2011,Marini_PRA_2011}, which is neglected in the averaging approach \cite{Agrawal2001}. Note that while $\Upsilon_{Si}$ is a real quantity, $\Upsilon_{Au}$ is complex and accounts for the nonlinear loss of gold. The nonlinear parameters of silica ($\Upsilon_{Si}$) and gold ($\Upsilon_{Au}$) are plotted as functions of the carrier wavelength $\lambda_0$ in Figs. \ref{Kerr_NL_COEFF_Fig},\ref{Gold_NL_COEFF_Fig_m_both}. In both figures, the full and dashed curves represent the $m=0$ and $m=1$ modes, while blue and green colors correspond to wire radii $r=50,250 ~nm$. Note that the real part of the gold nonlinear parameter is much greater than the Kerr nonlinear parameter of silica in the spectral region considered. Also, if $r\ll\lambda$, the nonlinear parameters of the $m=1$ mode are much smaller than those for the $m=0$ mode since they are much less confined. In this limit, as discussed in the previous section, while the fundamental $m=0$ mode is tightly confined to the metal surface and propagates only for a few wavelengths, the $m=1$ mode is much less localized and can propagate for longer distances (long-range guided SPP mode). This reduction in loss is at the cost of a weaker effective nonlinearity.

\begin{figure}
\centering
\begin{center}
\includegraphics[width=0.45\textwidth]{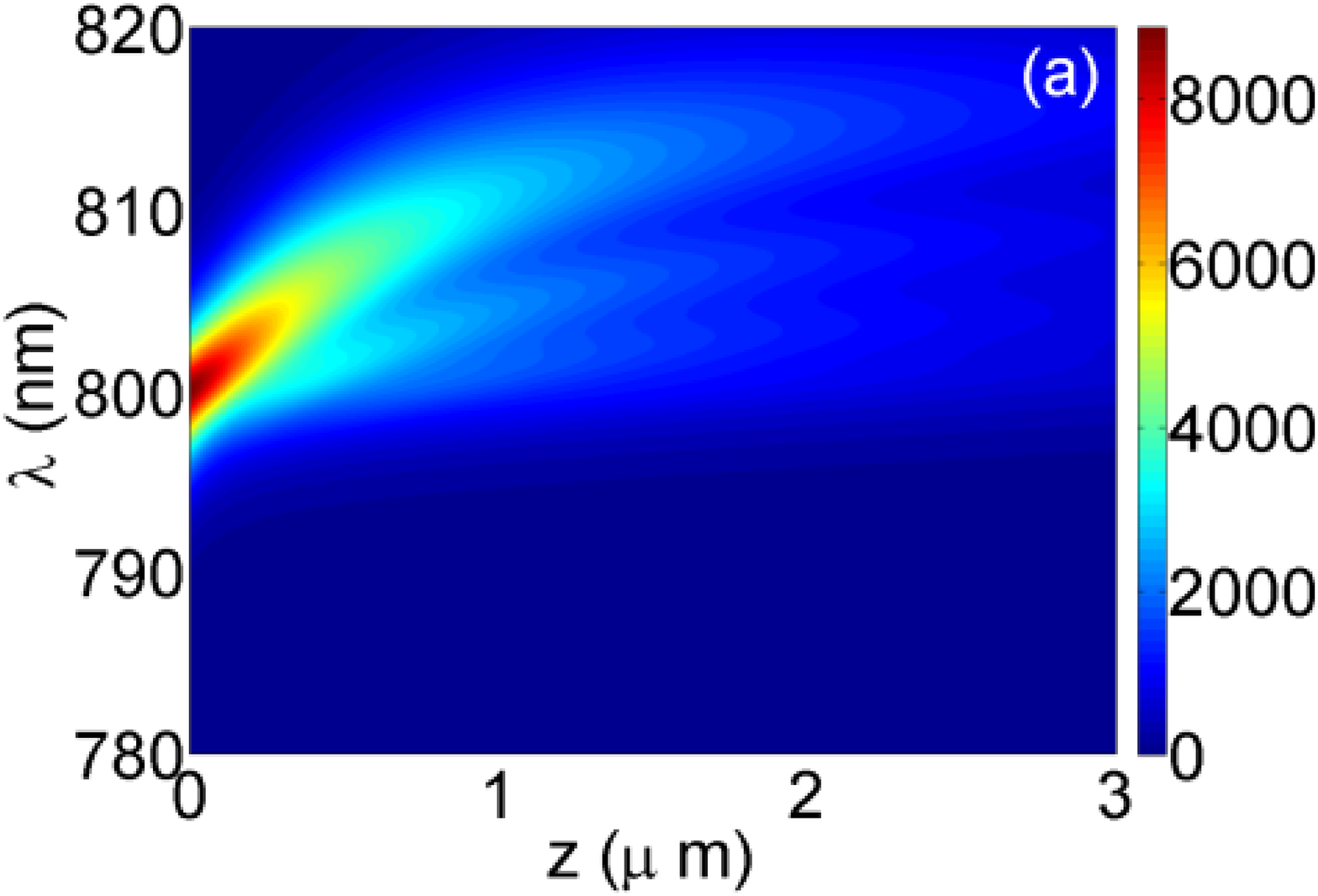}
\includegraphics[width=0.45\textwidth]{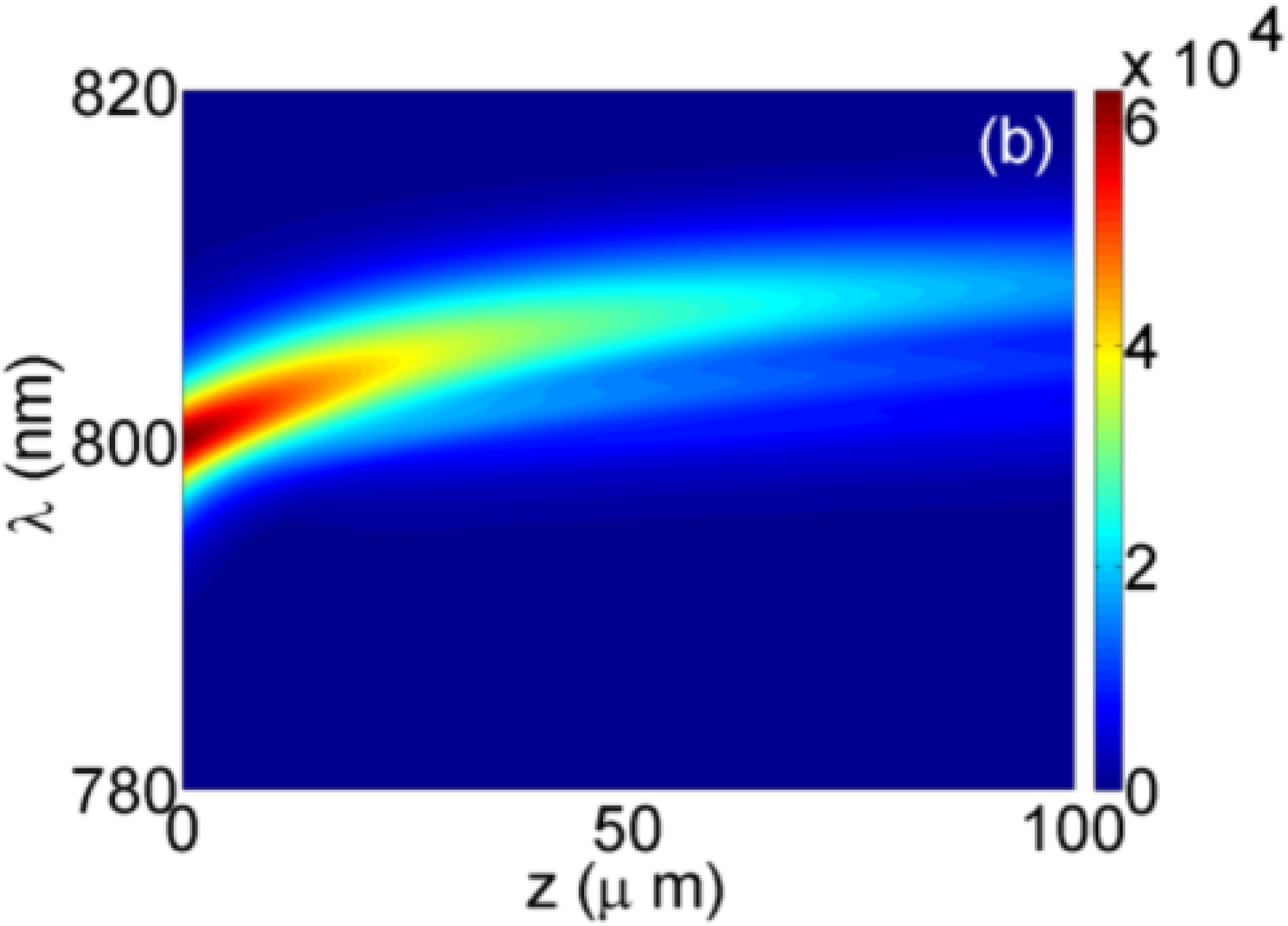}
\caption{Nonlinear propagation of an optical pulse along a gold nanowire with radius $r= 50 nm$ surrounded by silica glass for: (a) $m=0$ and an instantaneous input power $P_{in} = 1 \times 10^4 W$; (b) $m=1$ and an instantaneous input power $P_{in} = 5.3 \times 10^5 W$. In both figures the input pulse is a hyperbolic secant $\psi(0,t)=\sqrt{P_{in}}{\mathrm sech}(t/t_0)$, with $t_0=106 fs$. The contour-plot displays the modulus of the Fourier transform of the optical amplitude: $|\psi(z,\omega)|$. }
\label{Cntrplt_Fig}
\end{center}
\end{figure}

The formulation of the propagation equation Eq. (\ref{PropEq}) constitutes the main result of this paper. We have numerically solved Eq. (\ref{PropEq}) using the fast Fourier split-step algorithm \cite{Agrawal2001}. As we have already shown, when $\lambda_0\approx800nm$ the Kerr nonlinearity of silica does not play a significant role and can be neglected. For the numerical simulations we have considered a hyperbolic secant input pulse: $\psi(0,t)=\sqrt{P_{in}}{\mathrm sech}(t/t_0)$, with $t_0=106 fs$. $P_{in}$ is the instantaneous pulse power, which can be directly calculated from the average power of the laser source: $P_{in}=C_{eff}P_{av}/(2\nu_{rep}t_0)$, where $C_{eff}$ is the launch efficiency of the laser beam into the gold nanowire, $\nu_{rep}$ is the repetition rate and $2t_0$ is the pulse duration. The instantaneous power is kept well below the damage threshold power of gold, above which melting, ablation and vaporization occur ($P_{dam} \approx 10^6 W$) \cite{Stuart_JOSAB_1996}. In Fig. \ref{Cntrplt_Fig}, the numerical propagation along a gold nanowire with radius $r= 50 nm$ surrounded by silica glass is depicted for (a) $m=0$,  $P_{in} = 1 \times 10^4 W$ and (b) $m=1$, $P_{in} = 5.3 \times 10^5 W$. In this contour-plot the modulus of the Fourier transform of the optical amplitude ($|\psi(z,\omega)|$) is shown. The $m=0$ TM mode is highly nonlinear and significant nonlinear dynamics can be observed even for relatively small optical power. However, such a high nonlinearity is paid for by high loss, limiting the effective propagation length to $L \approx 2 \mu m$ (see Fig. \ref{Cntrplt_Fig}a). The hybrid polarized $m=1$ mode supports long-range guided SPP modes so that both the nonlinear and loss coefficients are significantly smaller. In this case, in order to observe a strong nonlinear dynamics, it is necessary to use a considerably higher optical power. For the $m=0,1$ modes a signature red-shift indicates the presence of a thermo-modulational interband nonlinearity, analogously with the Raman effect \cite{Blow_IEEE_1989}.

\begin{figure}
\centering
\begin{center}
\includegraphics[width=0.45\textwidth]{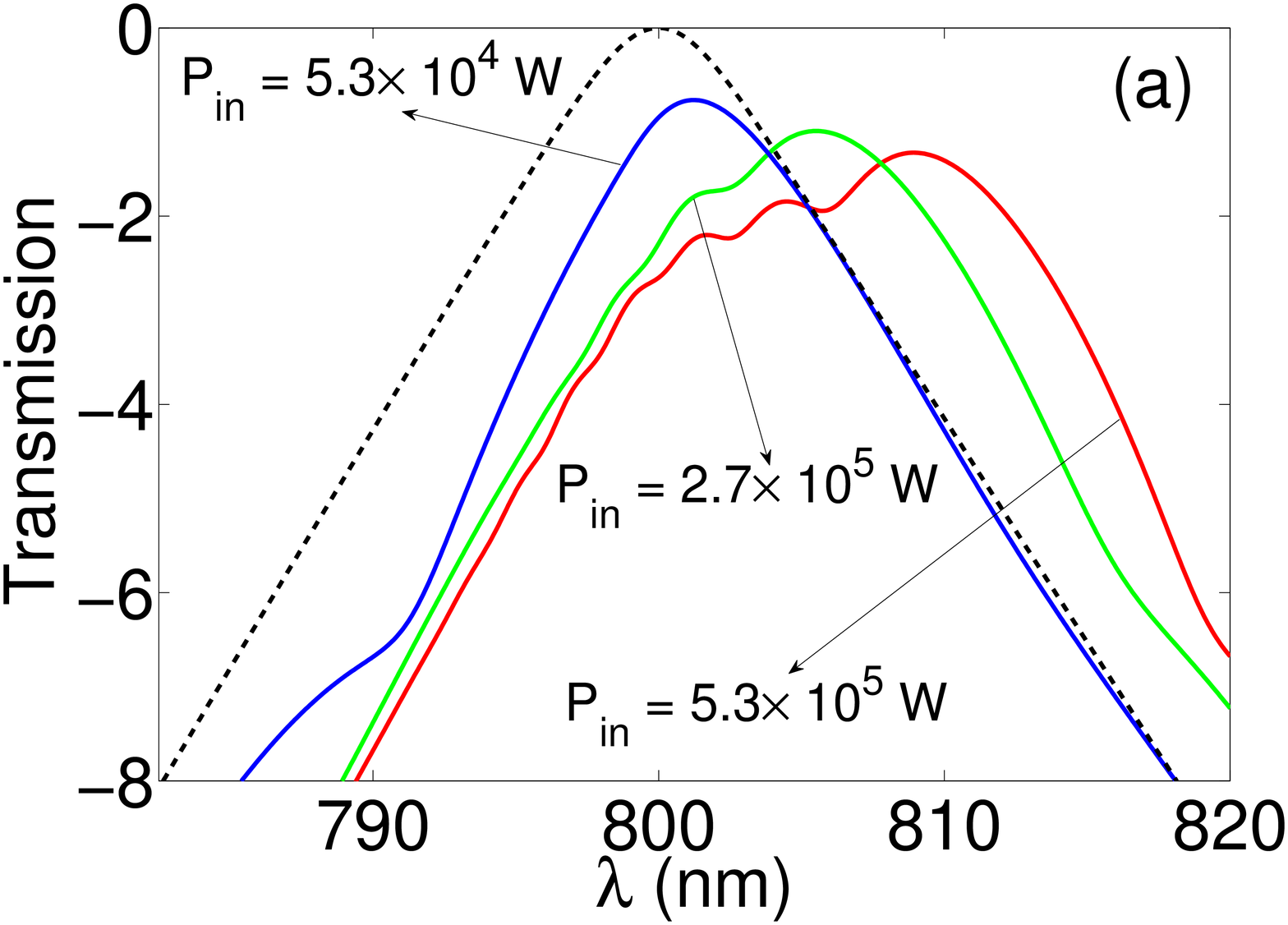}
\includegraphics[width=0.45\textwidth]{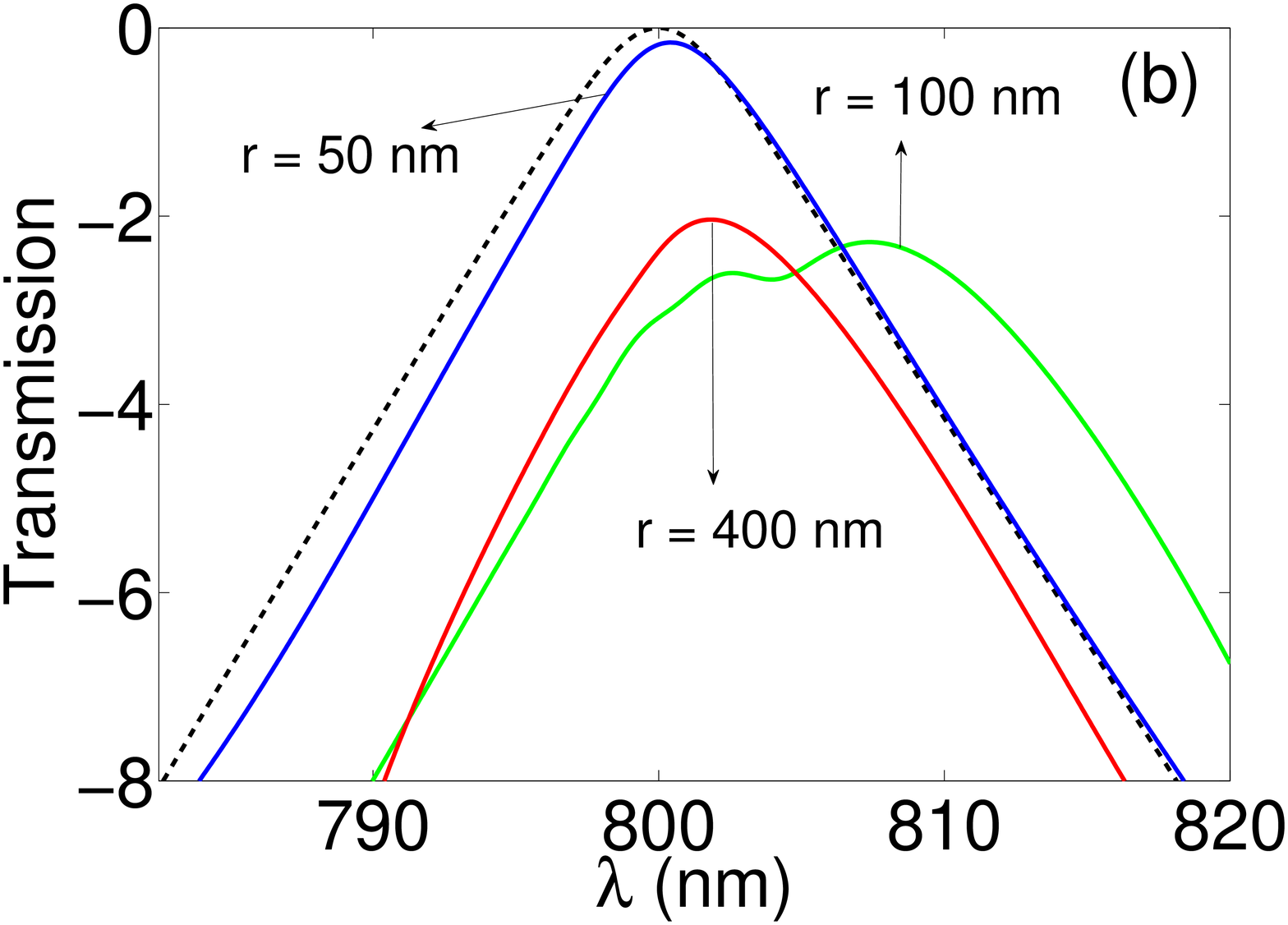}
\caption{Transmission spectrum (${\cal T} = ln|\psi(L,\omega)/\psi_{0M}|$, where $\psi_{0M}={\mathrm max}[\psi(0,\omega)]$) of the $m=1$ long-range mode. (a) ${\cal T}$ is calculated at a propagation length of $L = 100 \mu m$, for a fixed wire radius of $r=50nm$ and for different input optical powers: $P_{in} = 5.3 \times 10^4 W$ (blue curve), $P_{in} = 2.7 \times 10^5 W$ (green curve) and $P_{in} = 5.3 \times 10^5 W$ (red curve). (b) Transmission spectrum (${\cal T}$) for a propagation length of $L = 20 \mu m$, fixed input power $P_{in} = 5.3 \times 10^4 W$ and for different nanowire radii: $r = 50 nm$ (blue curve), $r = 100 nm$ (green curve) and $r = 400 nm$ (red curve). The black dashed curves represent the input spectrum ($ln|\psi(0,\omega)/\psi_{0M}|$).}
\label{Transmission_Fig}
\end{center}
\end{figure}

This red-shift is the natural consequence of the intrinsic delayed mechanism governing the thermo-modulational interband nonlinear susceptibility of gold. In the time domain the frequency red-shift is accompanied by a small pulse delay of  order $\approx 1 fs$. We emphasize that neither the Kerr nor the Raman nonlinearities of silica are large enough to produce the reported red-shift for the propagation lengths considered. The strong red-shift is accompanied by a large time-delayed nonlinear loss, as can be understood from Figs. \ref{Transmission_Fig}a,b. 

In Fig. \ref{Transmission_Fig}a, the transmission spectrum (${\cal T} = ln|\psi(L,\omega)/\psi_{0M}|$, where $\psi_{0M}={\mathrm max}[\psi(0,\omega)]$ and $L = 100 \mu m$) of the $m=1$ long-range mode of a gold nanowire with radius $r = 50 nm$ is depicted for several input powers: $P_{in} = 5.3 \times 10^4 W$ (blue curve), $P_{in} = 2.7 \times 10^5 W$ (green curve) and $P_{in} = 5.3 \times 10^5 W$ (red curve). The black dashed curve corresponds to the normalized input spectrum on a logarithmic scale ($ln|\psi(0,\omega)/\psi_{0M}|$). Note that as the input power increases the red-shift increases and the transmission peak decreases accordingly as a consequence of nonlinear loss. Also, as the power increases, the transmission spectrum displays some weak oscillations, which resemble Kerr-related self-phase modulation. Indeed, the dispersion length is much longer than the nonlinear length so that only the linear loss affects the nonlinear dynamics.

\begin{figure}
\centering
\begin{center}
\includegraphics[width=0.45\textwidth]{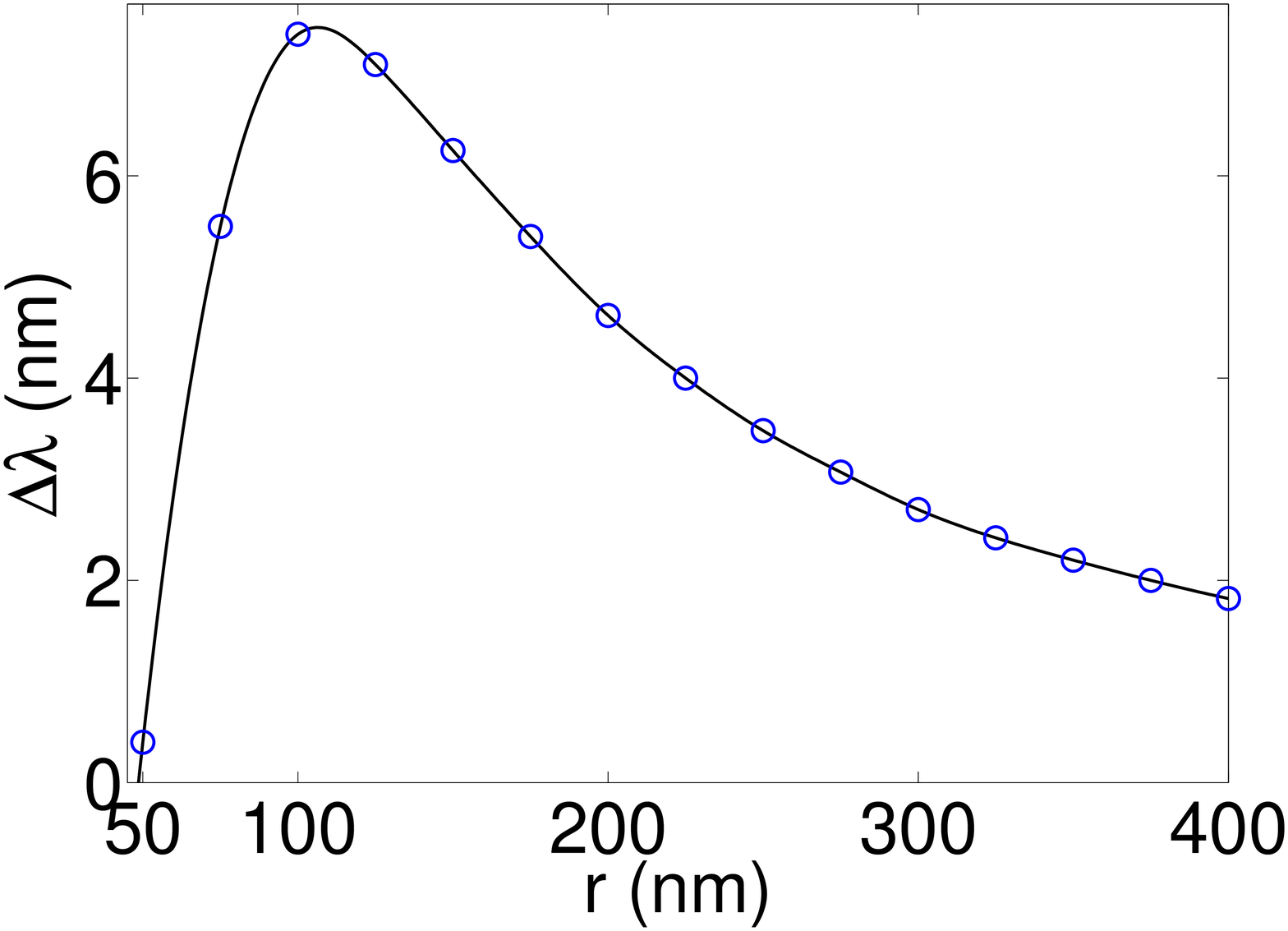}
\caption{Wavelength red-shift $\Delta\lambda$ as a function of the wire radius $r$ for the $m=1$ long-range mode, fixed input power $P_{in}=5.3 \times 10^4 W$ and the propagation length $L = 20 \mu m$.}
\label{Shift_Fig}
\end{center}
\end{figure}

In Fig. \ref{Transmission_Fig}b, the transmission spectrum (${\cal T} = ln|\psi(L,\omega)/\psi_{0M}|$, where $L = 20 \mu m$) of the $m=1$ long-range mode is shown for several nanowire radii: $r = 50 nm$ (blue curve), $r = 100 nm$ (green curve) and $r = 400 nm$ (red curve). The black dashed curve represents the input spectrum ($ln|\psi(0,\omega)/\psi_{0M}|$) and the input power is fixed to $P_{in} = 5.3 \times 10^4 W$. Note that the linear properties of the $m=1$ mode are non-trivial (see Fig. \ref{Beta_Im_R_TOT_m_1_Fig}) and, as a consequence, the power-dependence of the red-shift and the transmission peak is non-monotonic. This means that an optimal radius $r = r_{o}$ exists, where the achievable red-shift is maximum. In Fig. \ref{Shift_Fig}, the red-shift of the transmission peak of the $m=1$ long-range mode is plotted as a function of wire radius, for fixed input power $P_{in}=5.3 \times 10^4 W$ and for a propagation distance $L = 20 \mu m$. The open blue circles represent the results of numerical simulations, while the black curve corresponds to an interpolation of the numerical results. The maximum red-shift attainable at this input power is $\Delta\lambda \approx 7 nm$ for a radius of $r_{o}\approx 110 nm$. The dependence of the thermo-modulational interband nonlinear coefficient ($\Upsilon_{Au}$) and the absorption coefficient ($\alpha=2\beta''_0$) on $r$ strongly affects the nonlinear dynamics and as a consequence the red-shift attainable, which reaches a maximum at $r=r_o$. If the input power increases, the maximum red-shift increases accordingly. 

\section{Summary}

In this paper we have described the thermo-modulational interband nonlinearity of gold starting from its band structure. Electrons in the conduction band are heated by an ultrashort optical pulse and the interband dielectric properties are modulated accordingly. Using a semiclassical approach, we have calculated the imaginary part of the dielectric constant of gold, accounting for interband absorption, which basically depends on the joint density of states. In turn, we have been able to describe the temperature dependence of the dielectric susceptibility of gold and have modeled the heating and the relaxation of the conduction electrons using a two-temperature model. We have discovered that the metal nonlinearity is basically characterized by a delayed mechanism, similar to the Raman effect in some senses, but with a much longer response time ($\approx 300 fsec$). Also, in contrast to the Raman effect, the thermo-modulational interband susceptibility is complex-valued, providing a delayed nonlinear loss/gain. The optical propagation of surface plasmon polaritons is strongly affected by the metal nonlinearity, which we have found to be several orders of magnitude larger than the Kerr nonlinearity of fused silica. We have derived, for the first time to our knowledge, a generalized nonlinear Schr\"odinger equation suitable for modeling the optical propagation of SPPs along a gold nanowire surrounded by silica glass. Solving this equation using a fast Fourier split step algorithm, we have found that the signature of the thermo-modulational interband nonlinearity is a {\it red-shift} of the optical pulse. This red-shift results from the intrinsic time-delayed nature of the thermo-modulational interband nonlinearity of gold. We believe that this novel nonlinear effect may be important for frequency conversion in plasmonic devices. We have also provided some details on the expected red-shift, its dependence on the wire radius and the optical power necessary to observe it experimentally.

\acknowledgements
This research is supported by German the Max Planck Society for the Advancement of Science (MPG).

\end{document}